\documentclass[12pt, prd]{revtex4}
\usepackage{amsmath}
\usepackage{amssymb}
\usepackage[utf8]{inputenc}
\usepackage{graphicx,epsfig}
\usepackage{xcolor}

\begin{document}

\title{
Shadow of the Moon and general relativity:
Einstein, Dyson, Eddington and the 1919 light deflection
}
\author{Jos\'{e} P. S. Lemos}
\affiliation{Centro de de Astrof\'isica e Gravita\c c\~ao - CENTRA,
Departamento de F\'{\i}sica, Instituto Superior T\'ecnico - IST,
Universidade de Lisboa - UL, Avenida Rovisco Pais 1, 1049-001,
Portugal\\ email: joselemos@ist.utl.pt}

\begin{abstract}

The eclipse of the Sun of 1919 was fundamental in the development of
physics and earns a high place in the history of science. Several
players took part in this adventure. The most important are Einstein,
Dyson, Eddington, the Sun, the Moon, Sobral, and Principe.  Einstein's
theory of gravitation, general relativity, had the prediction that the
gravitational field of the Sun deflects an incoming light ray from a
background star on its way to Earth.  The calculation gave that the
shift in the star's position was 1.75 arcseconds for light rays
passing at the Sun's rim.  So to test it definitely it was necessary
to be in the right places on May 29, 1919, the day of the eclipse.
That indeed happened, with a Royal Greenwich Observatory team composed
of Crommelin and Davidson that went to Sobral, and that was led at a
distance by the Astronomer Royal Frank Dyson, and with Eddington of
Cambridge University that went to Principe with his assistant
Cottingham.  The adventure is fascinating, from the preparations, to
the day of the eclipse, the data analysis, the results, and the
history that has been made. It confirmed general relativity, and
marked an epoch that helped in delineating science in the post eclipse
era up to now and into the future.  This year of 2019 we are
celebrating this enormous breakthrough.


\end{abstract}

\maketitle

\newpage

\section{Introduction}

It was written in the stars that the May 29, 1919, eclipse of the Sun
would be the most important eclipse in the history of humankind.

In a total eclipse of the Sun, the Moon blocks the sunlight and casts
a shadow on the Earth, turning the sky dark and letting the stars
shine like it was night, the whole scene yielding back to the normal
daylight as the required alignment is slowly undone with
the turning of the celestial sphere at the place.
The totality, i.e., the period during which an eclipse is total,
takes about 5
minutes.

In ancient times the phenomenon was not understood in rational terms,
only after the emergence of the recent civilizations, like the
Babylonian and Assyrian, eclipses started to be understood as natural
phenomena explained by an alignment between the Sun, Moon and Earth.
The Earth was fixed, the celestial sphere with its stars turned around
day after day, the Moon moved in the celestial sphere with a period of
a month, and the Sun moved along the ecliptic, the great circle on the
celestial sphere representing the Sun's path during the year.

With the advent of the Copernican revolution, that the Sun is at the
center and the planets move around it, the understanding of the
movements in the celestial sphere became highly simplified. One could
now explain the motion of the Sun along the ecliptic through the
motion of the Earth around the Sun. As the Earth moves the projection
of the Sun into the celestial sphere changes, returning to the same
point after tracing the ecliptic during one year.  The Moon moves
around the Earth during one month with an orbit slightly inclined
relatively to the plane of the Earth's orbit. So, when it occurs that
the Moon is new and the Moon's path is crossing the plane of the
Earth's path there is an eclipse of the Sun, in case the cross is with
a full Moon the eclipse is of the Moon, and this is the reason for the
name ecliptic, the path where eclipses occur.

With the emergence of Newton's gravitation and precise celestial
mechanics, eclipses and their places could be predicted with
ease. With the development of telescopes through the use of photography
and spectroscopy and the rise of astrophysics, it was possible to make
observations of the Sun's cromosphere and
corona during eclipses helping thus to
determine its composition and physical properties.

But suddenly, solar eclipses started to have a new significance. They
could be used to prove or disprove fundamental physics. Indeed,
general relativity was completed by 1915, and Einstein, its author,
predicted that the gravitational field of the Sun should deflect an
incoming light ray from a background star on its way to Earth, such
that the shift in the star's position was 1.75 arcseconds for light
rays passing at the Sun's rim.  If correct, one would know that the
world is governed by general relativity, a theory of gravitation that
connects space, time, and matter at a fundamental level. To confirm
the light deflection prediction of general relativity one needed a
solar eclipse such that the stars near the Sun's rim could be seen and
their displacements measured.

In May 29, 1919, there was an eclipse with several bright stars
in the background near the Sun giving the perfect conditions
to measure the deflection effect, the more bright stars
one has the better the results can be trusted.
Due to the narrow bandwidth shadow that the Moon projects, solar
eclipses usually are in remote parts of the world, so preparations to
transport the telescopes and have the correct equipment and provisions
took months. On the day of the eclipse all had to be ready so that
during the five minutes of totality observations and photographic
plates would work perfectly and give good results.
If it rained or the sky were filled with heavy clouds all
the work would have been for nothing.
The 1919 eclipse would be no exception, one would
have to go to Sobral, north of Brazil, and
Principe, a Portuguese island at the time off the
coast of west Africa.

British astronomers and astrophysicists decided it was time to test
general relativity. Dyson, the Astronomer Royal, would collaborate
with 
Crommelin and Davidson that would make the observations in Sobral, and
Eddington would go with Cottingham to Principe.  The enterprise was a
success, with Einstein's general relativity being confirmed when the
announcement of the results was made in London, in a
joint meeting of the Royal Society and Royal Astronomical Society, on
November 6, 1919.

This year of 2019 we celebrate one hundred years of this adventure.
It is worth to register the seven main protagonists that took part in
it, to understand the principles that lead to light deflection in
general relativity, and to go through the 1919 eclipse in detail,
namely, the preparations, the day of the eclipse, the data analysis, the results,
and how history looks back at it.  In confirming general relativity
the 1919 eclipse science accomplishment outlined in some way the
post eclipse era up to now and into the future.
So let us wander
through the 1919 eclipse wonders.

\section{The protagonists}

\subsection{Einstein}

\begin{figure}[h]
\centering 
{\includegraphics[scale=0.1]{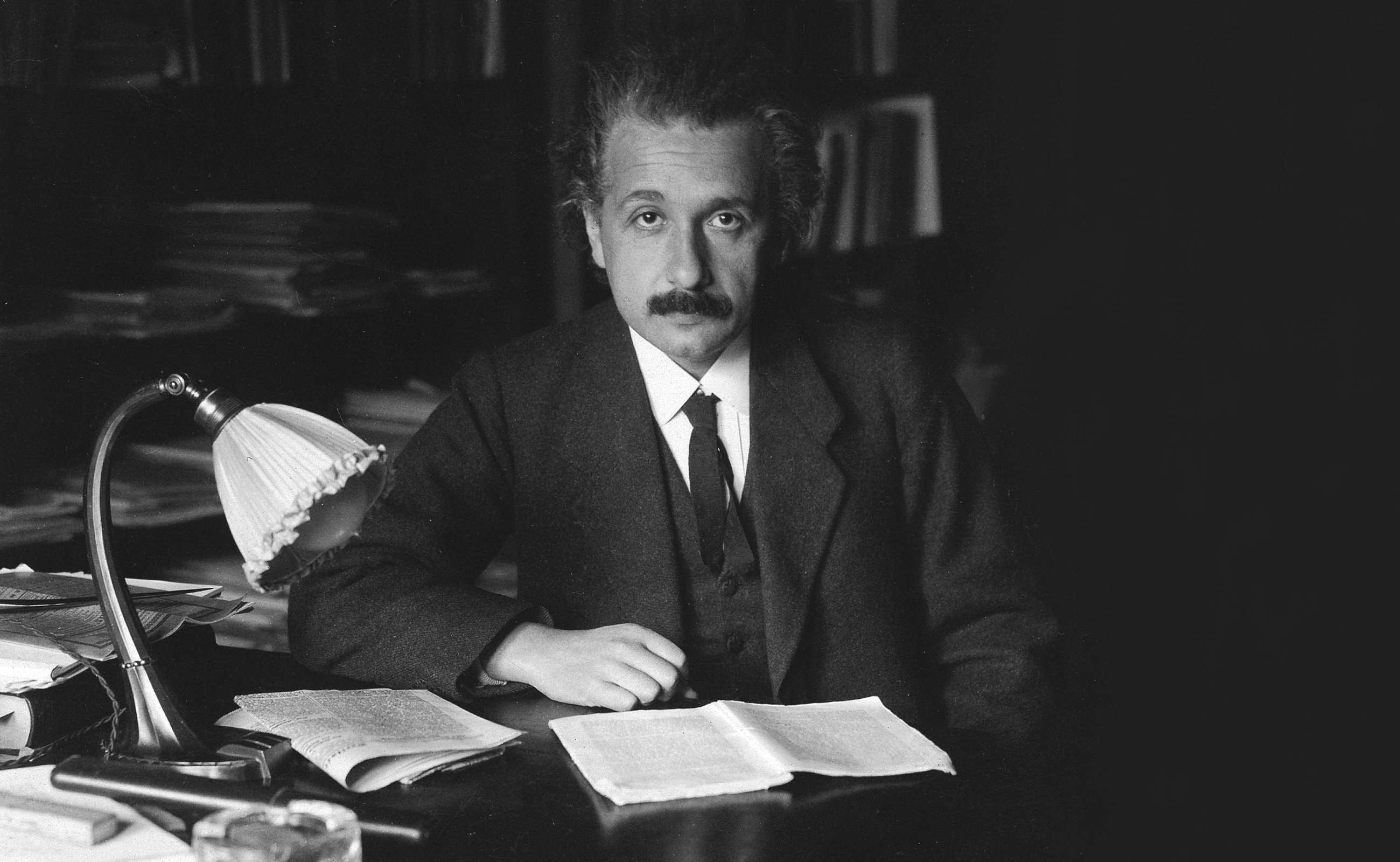}}
\caption{Einstein in Berlin around 1919.}
\label{Einstein around 1919 in Berlin}
\end{figure}

Einstein, see Fig.~\ref{Einstein around 1919 in Berlin},
had a very important role in a number of areas in physics but the
pinnacle is without a doubt the creation of general relativity. After
great persistence from a work that started with the idea of the
principle of equivalence in 1907 \cite{einstein1907,einstein1911}
and continued through the collaboration with Grossmann
in the entwurf theory of gravitation
\cite{einsteingrossmann1914,einstein1914aentwurfdeflection,einstein1914b},
Einstein
presented to the Prussian Academy of Sciences, in November 18 and 25, 1915
\cite{einstein1915perihelion,einstein1915}, see also \cite{einstein1916},
a totally new
theory, namely, a covariant, tensorial, and relativistic theory, that
he immediately called the general theory of relativity, or simply,
general relativity.
Einstein
equation that governs general relativity is
\begin{equation}
G_{ab}=\frac{8\,\pi\,G}{c^4}\,T_{ab}\,, 
\end{equation}
where $G_{ab}$ is a quantity that represents the geometry of spacetime
called the Einstein tensor, $T_{ab}$ is a quantity that represents the
matter content of the spacetime called the energy-momentum tensor,
$G$ is the constant of gravitation, and $c$ is the speed of light,
see \cite{renn} for the genesis of this equation. In
a stroke, the theory confirms the Minkowskian spacetime notion, states
that gravitation is geometry, spacetime is curved, and particles follow
geodesics. For
accounts of this period in Einstein's life
see
\cite{clarkebio,pais1,pais2}.

In its more than one hundred years, general relativity has passed
through very rigorous tests, it is accepted as the standard theory of
gravitation, and is considered one of the great feats in history.
Notwithstanding all these achievements, gravitation is the most
intriguing of all the known interactions.

The tests and implications of general
relativity are many and profound. 
Weak field classical tests
within the solar system are
the perihelion precession of Mercury,
the light deflection in the gravitational field
of the Sun, the gravitational redshift
Doppler effect,
and the Shapiro gravitational time delay in the radar echo.
Technological applications of general
relativity, are now current, as the
global position system, or GPS, would not work
at all without the  general relativistic
corrections related to  
the gravitational redshift
Doppler effect,
necessary to synchronize clocks
in the satellites
with clocks on the Earth's surface. 
Gravitational lensing is an
abundant special case
of light deflection and 
of great importance to understand the gravitational
mass and gravitational structure of the
Universe.
Cosmology, the dynamical and physical
study of the Universe, was started by
Einstein in 1917
with a static finite universe, continued with the proposal by
Friedmann, Lemaitre, and Hubble for an expanding universe, along
with the establishment of the big bang scenario through the
discovery of the cosmic microwave background radiation, up to the
establishment of the acceleration of the Universe, and to the most
recent astonishing developments, that converged
in the awarding of the shared
2019 Nobel Prize in Physics 
to Peebles of Princeton University,
one of the exponents in the field
throughout the last six decades.
Fundamental theories, theories that
make the unification of gravitation and
electromagnetism, were initiated by Weyl in 1918,
and continued by Eddington and Einstein. 
Now they are called theories of everything
and try to unify the four fundamental fields
in a unique quantum scheme.
Black holes, the geometric object par excellence in
general relativity, were found  by Oppenheimer and Snyder in 1939
as the endpoint of gravitational collapse
and thus occurring necessarily in nature.
Millions of solar mass
black holes float through our galaxy, 
and all, or almost all, galaxies contain a
central supermassive black hole in its center.
Gravitational waves, spacetime waves predicted by Einstein in
1916, were detected indirectly in the binary pulsar
discovered by Hulse and Taylor in 1976,
which gave the 
Nobel prize in 1993,
and detected  directly in 2015 by the LIGO
antennas, from the collision
of two black holes, which in turn gave the 
Nobel prize in 2017.
General relativity has left an immense and
amazing legacy and we are still
in the middle of many of its developments.

\subsection{Dyson}

\begin{figure}[h]
{\includegraphics[scale=0.68]{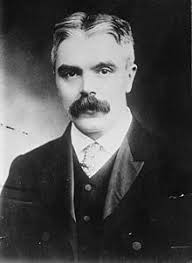}}
\caption{Dyson around 1910.}
\label{Dyson}
\end{figure}

Frank Dyson, see Fig.~\ref{Dyson},
excelled in astronomy, was Astronomer Royal and 
Director of the Royal Greenwich Observatory.
He had been a Cambridge
student in Trinity College
scoring Second Wrangler in 1889, was elected
 Fellow of the Royal Society, i.e., FRS,
in 1901, and 
worked on the Astrographic Catalogue
published in 1905.

Dyson was a world specialist in solar eclipses
and an expert in the solar corona and chromosphere.
He was present in six eclipses with success in all,
in particular,
his first eclipse was in Ovar, Portugal, for
the May 28, 1900, eclipse, and was 
a member of the Joint Permanent Eclipse Committee,
which was founded in 1884 by the British
to strengthen expertise on the subject
\cite{eddingtonondyson1940}.

Through Eddington, he understood the importance and possibilities of
the 1919 eclipse that could prove or disprove general relativity and
invited Eddington himself in turn to participate in the observations.
After all the limelight had passed he knew that the eclipses' findings
would be recorded as Eddington's.

\begin{figure}[h]
{\includegraphics[scale=1.98]{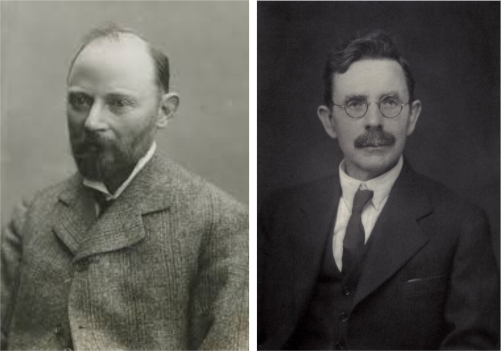}}
\caption{Left: Crommelin around 1919. Right: Davidson around 1919.}
\label{crommelindavidson}
\end{figure}

Two Dyson's assistants at the Royal Greenwich Observatory participated
in this adventure and were sent to Sobral to do the eclipse
observations.  Andrew
Crommelin, see Fig.~\ref{crommelindavidson},
Irish astronomer specialist in comets, led the
expedition, and on the day was in charge of the
4 inch telescope that yielded the important data in the end. He
substituted in the last minute Aloysius Cortie, a Jesuit astronomer
that did not travel because of work, possibly he had to take immediate
charge of being Director of the Stonyhurst College Observatory.
Crommelin was president of the Royal Astronomical Society from 1929 to
1931.
Charles Davidson, see Fig.~\ref{crommelindavidson},
was a computer, i.e., did
arithmetic calculations for problems in astronomy at
the Royal Greenwich Observatory, and scaled up the hierarchy through
his skills with instruments and telescopes.
He accompanied Dyson and Eddington in other eclipses.
He was elected FRS in 1931, showing his
great competence. In Sobral, he operated the 13 inch astrographic
telescope that did not work well.

\subsection{Eddington}

\begin{figure}[h]
{\includegraphics[scale=0.65]{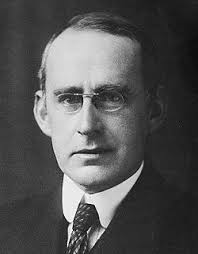}}
\caption{Eddington around 1915.}
\label{Eddington around 1915}
\end{figure}

Eddington, see Fig.~\ref{Eddington around 1915},
professor in the University of Cambridge,
shined in astrophysics and gravitation.
He was the first and only second-year student
that got the much aimed Senior 
Wrangler, i.e., First Wrangler.
Was elected FRS in 1914, President of the Royal Astronomical Society
in the period
1921-1923, and President of IAU in the period 1938-1944 \cite{allie},
see also \cite{eddbiographykillmister,eddbiographystanley}.

Eddington was the first to understand the physics of the stars at its
center with temperatures of 10 million degrees.  In 1918 he introduced
general relativity to the British audience which were still grasping
to understand special relativity.  He wrote several outstanding books
on the theory, read with enthusiasm worldwide for many generations up
to now.  He was the first to give the correct quadrupole formula for
gravitational radiation as a 1/2 term was missing in Einstein's
deduction, he introduced a parameterization scheme for general
relativity through a post-Newtonian approximation which appeared in
his book in mathematical relativity and was much later developed into
the parameterized post-Newtonian, or PPN, formalism, he initiated the
study of gravitational radiation emission using a rotating star
mimicked by a rod, and he examined the problem of $n$ bodies in
general relativity with interesting results.  He was lured into
cosmology and tried frantically a theory of everything without
success.  He was the main name in the 1919 expeditions that detected
the light deflection by the gravitational field of the Sun, he himself
having made the observations in Principe.  His many books for the
public understanding of science are fascinating.

\begin{figure}[h]
{\includegraphics[scale=0.4]{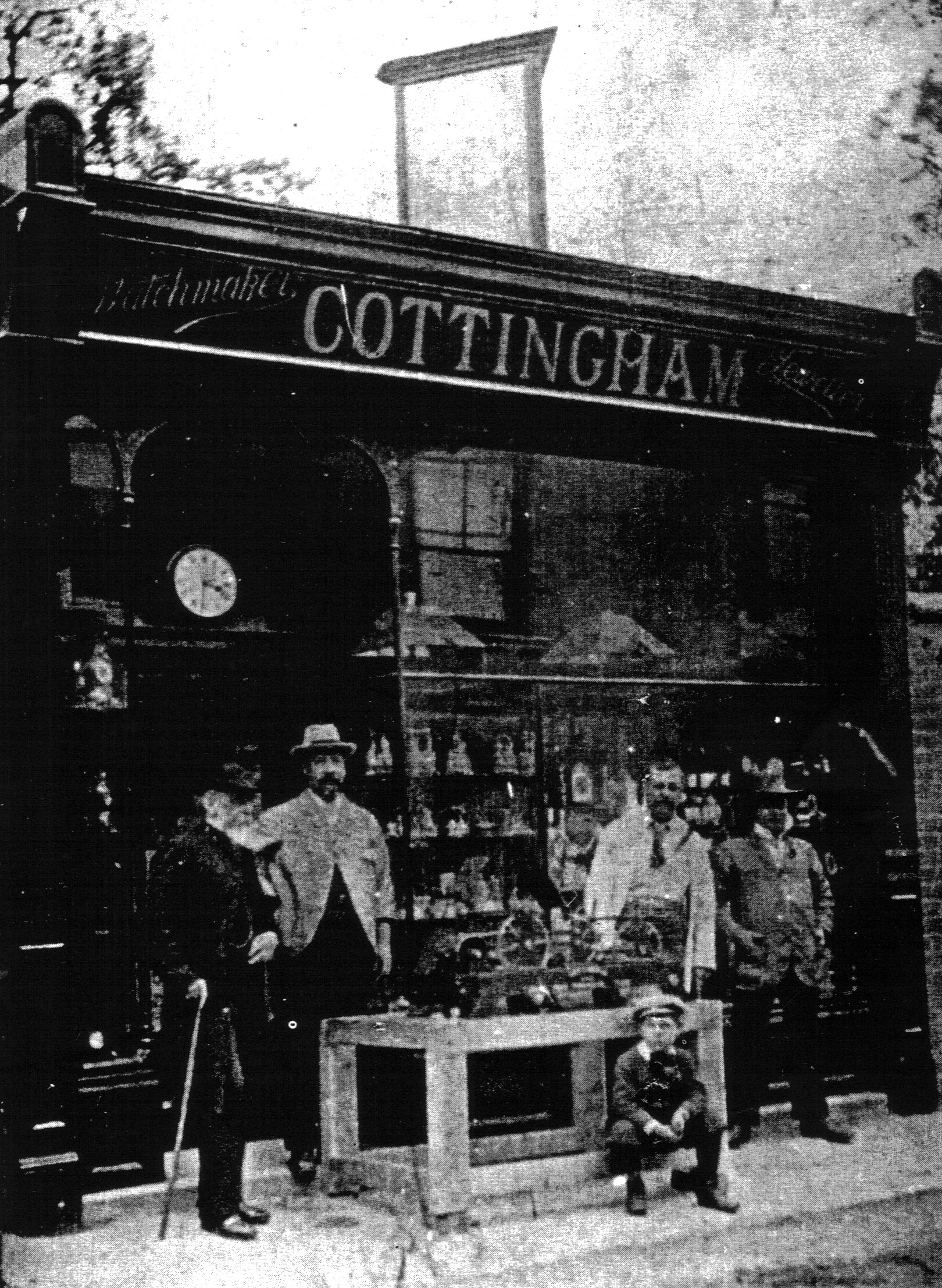}}
\caption{Cottingham, second from the left, in his watchmaker shop,
around 1910.}
\label{cottinghamshop}
\end{figure}

Edwin Cottingham, see Fig.~\ref{cottinghamshop},
made fame and money through 
the making of clocks as he was a 
gifted mechanical and electric technician. His shop
was in Thrapston, between Northampton and Cambridge, and
Cambridge University would frequently called him to help
in the maintenance of clocks and instruments. 
In Principe he helped 
Eddington in the mounting and maneuvering of the
telescope. 
Eddington narrates that, on the day
previous to the departure, in Greenwich,
he and Dyson were
discussing that Newtonian gravitation
gave half the amount of the deflection
predicted by Einstein's theory, and
that Eddington expected to observe
the full deflection. 
Cottingham was present and asked
``What will it mean if we get double the Einstein
deflection?''
which prompted a response from Dyson,
``Then Eddington will go mad, and you will
have to come home alone!'' \cite{eddingtononcottigham}, see
also \cite{allie}.

\subsection{The Sun}

\begin{figure}[h]
{\includegraphics[scale=0.17]{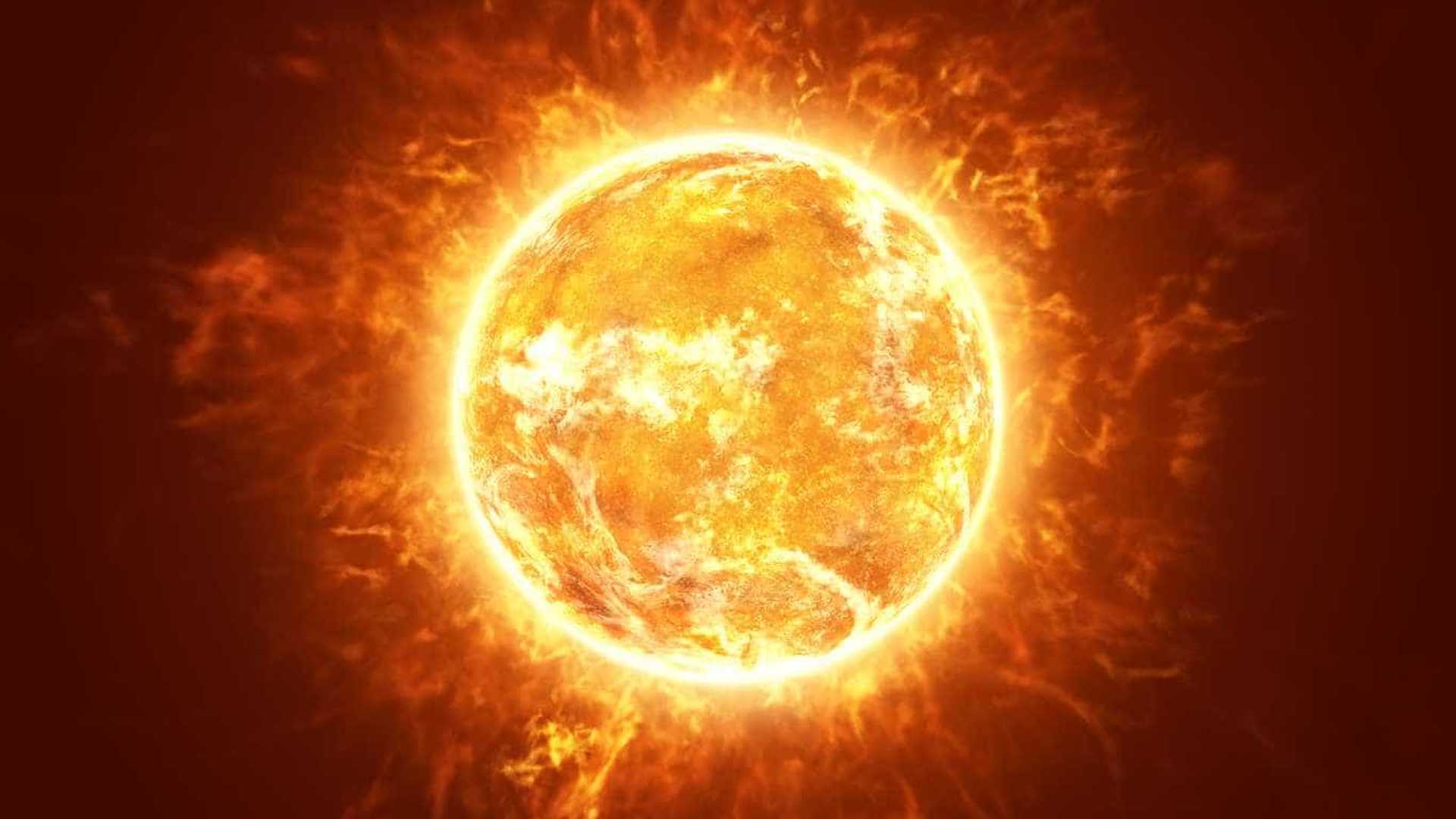}}
\caption{The Sun.}
\label{sol}
\end{figure}

The Sun, see Fig.~\ref{sol},
is our beloved star, it shines and deposits energy 
for terrestrial processes.
However, here
the interest is that the Sun
is source of gravitational field,
or in other words, it
distorts spacetime.

There is a dictionary for trading
words and concepts from
Newtonian gravitation to 
general relativity, like
gravitational force for spacetime
distortion, but it is not exhaustive
as the two theories 
have completely different ontologies
with general relativity being much more 
comprehensive.
We want to understand the motion of light
in a gravitational field like the Sun
through Newtonian gravitation and
through general relativity.

In Newtonian mechanics a particle moves according to Newton's second
law of motion $F=m\,\frac{d^2r}{dt^2}$, where $F$ is the applied
force, $m$ is the particle's inertial mass, $r$ is the particle's
position, $t$ is the time, and $\frac{d^2r}{dt^2}$ the particle's
acceleration.  Newtonian gravitation states that the force $F$ exerted
by the Sun in a particle is $F=\frac{GMm}{r^2}$, where $M$ and $m$ are
the Sun's and the particle's gravitational mass, respectively, $r$ the
distance between the Sun and the particle, and $G$ Newton's universal
constant, see, e.g., \cite{bargerolsson} for more on Newtonian
gravitation.  By experiment, it is known that inertial and
gravitational masses have the same value which for the particle we
have called $m$.  For the purposes of understanding the motion of
light, considered as a massless particle, note that in Newtonian
gravitation, the motion of a particle is generically governed by
$\frac{d^2r}{dt^2}=\frac{GM}{r^2}$, since inertial and gravitational
masses having the same value $m$, they cancel in the equation of
motion, i.e., the particle's mass $m$ does not enter into the problem.
So, light moving in a gravitational field can indeed be thought as
obeying Newtonian gravitation laws, the only 
hypothesis is that light
is a particle.  Whether the light particle has zero mass or any other
effective mass, it does not matter, since even the zero mass case can
be conceived as a particle of very tiny mass with that mass going to
zero in the limit. However, in the 19th century light started to be
considered as a wave, instead of a particle, in which case it would
not be influenced by gravitation, there would be no coupling between
light and gravitation, as it was thought.
\begin{figure}[h]
{\includegraphics[scale=0.8]{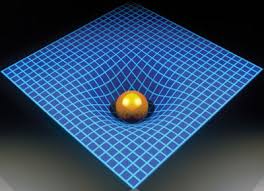}}
\caption{Spacetime distortion  by the Sun
here represented as space distortion
alone. The movement of massive
particles and massless particles such as
light is described as
geodesic motion in spacetime.}
\label{spacetime distortion}
\end{figure}
But even this idea of no
coupling between waves and gravitation had to be rethought within the
context of special relativity, as here light would have an effective
mass $m=E/c^2$ corresponding to 
the special relativistic energy $E$, and where $c$ is the velocity of light.
Since then in special
relativity a lightwave has an effective mass it would now
indeed couple to
gravitation and light, even as a wave,
 could follow Newton's laws.  So, the motion of light
in a gravitational field in Newtonian gravitation could not be settled
so easily, it could be the case that light is a wave that does not
couple to gravitation, or it could be the case that
light is a particle of
some mass, even zero mass, or a wave with effective mass $m=E/c^2$, in
both these cases light would couple and be influenced by the laws of
gravitation as described by Newton.

Einstein's general relativity, based on the principle of equivalence
that states that gravitational and inertial masses are the same thing,
arrives at the notion that spacetime is curved, in particular is
curved around the Sun, see Fig.~\ref{spacetime distortion}, and a
particle moves following this curvature.  General relativity
incorporates then naturally the motion of light within its spacetime
formalism and describes it as geodesic motion.

\subsection{The Moon}
\begin{figure}[h]
{\includegraphics[scale=0.85]{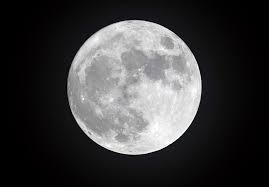}}
\caption{The full moon.}
\label{moon}
\end{figure}

The Moon, see Fig.~\ref{moon},
has a very important effect on Earth: The tides.
However, here the interest is that the Moon can put itself between the
Earth and the Sun yielding a full shadow of the Moon, i.e., a total
eclipse of the Sun.

\begin{figure}[h]
{\includegraphics[scale=1.60]{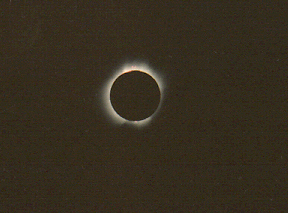}}
\caption{The July 11,
1991, shadow of the Moon from Playas del Coco, Costa Rica. This eclipse
also passed some hours later in Amazonia.}
\label{1991eclipsecostaricaamazonia}
\end{figure}

A total eclipse of the Sun is relatively rare.  The plane of the orbit
of the Moon around the Earth makes a 5$^{\rm o}$ angle with the
\begin{figure}[h]
{\includegraphics[scale=0.28]{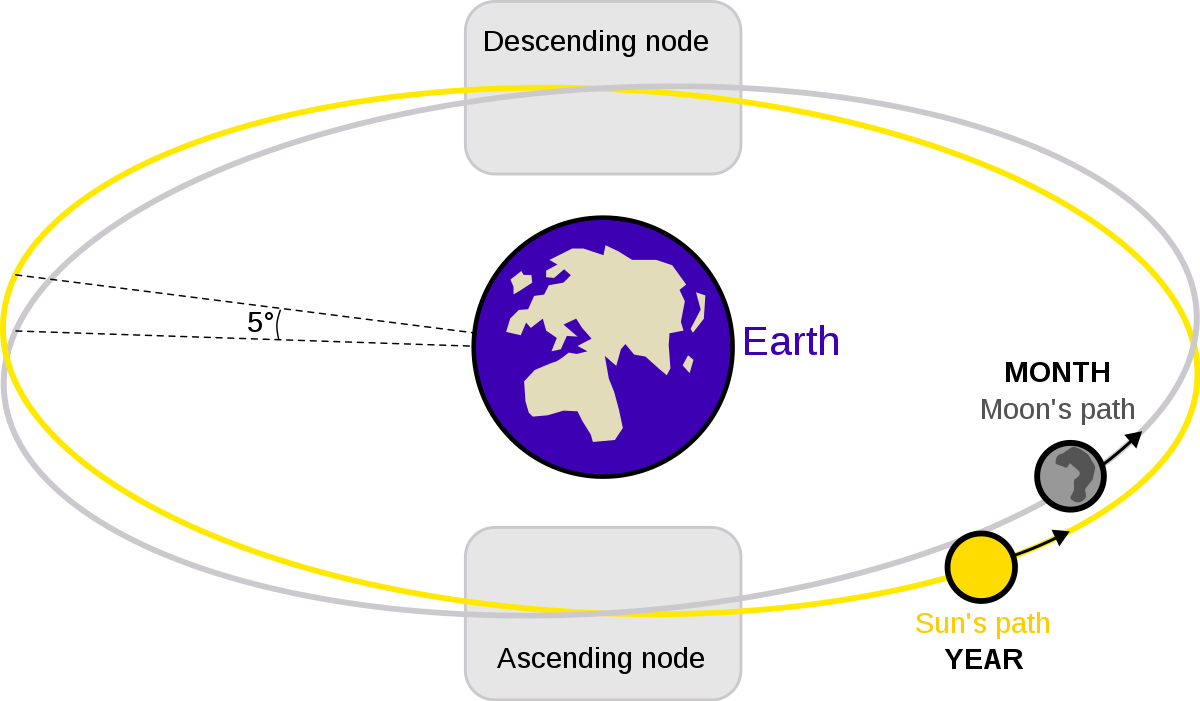}}
\caption{Lunar orbit showing the nodes when the orbit crosses the
ecliptic, the Sun's path on the
celestial sphere along the year.}
\label{nodeslunareclipsediagram}
\end{figure}
ecliptic, the plane of the projected annual orbit of the Sun around
the Earth onto the celestial sphere, see
Fig.~\ref{nodeslunareclipsediagram}.  This plane is of course the same
as the plane of the orbit of the Earth around the Sun.  The two orbits,
Moon's and Sun's,
cross at the nodes, one ascending the other descending.  When the Moon
is in between the Earth and the Sun one has a new moon. If the new
moon happens to be on a node then there is a total eclipse of the Sun,
At the eclipse, the disk of Moon
barely hides the disk of the Sun. If in the eclipse the Moon's orbit
is in the perigee, when it is closest to Earth, then the eclipse is
total, if the Moon's orbit is in the apogee, when it is
farthest from the Earth, then an annulus of the Sun can be seen and the
eclipse is annular.  When there is a total eclipse the shadow of the
Moon on Earth is about 200$\,$Km wide and can run a track roughly from
west to east of some 12 thousand Km, visiting countries and oceans. A
typical eclipse photograph is seen in
Fig.~\ref{1991eclipsecostaricaamazonia}.
For information on solar eclipses see, e.g.,
\cite{littmann,gren}.

\subsection{Sobral}

The 1919 eclipse would hit Earth along
several places, in particular
it would pass
through the north of
Brazil.  Henrique Morize, the director of Observatorio
Nacional in Rio de Janeiro, recommended to the British
Sobral 
as the right place to do the eclipse observations.
\begin{figure}[h]
{\includegraphics[scale=2.00]{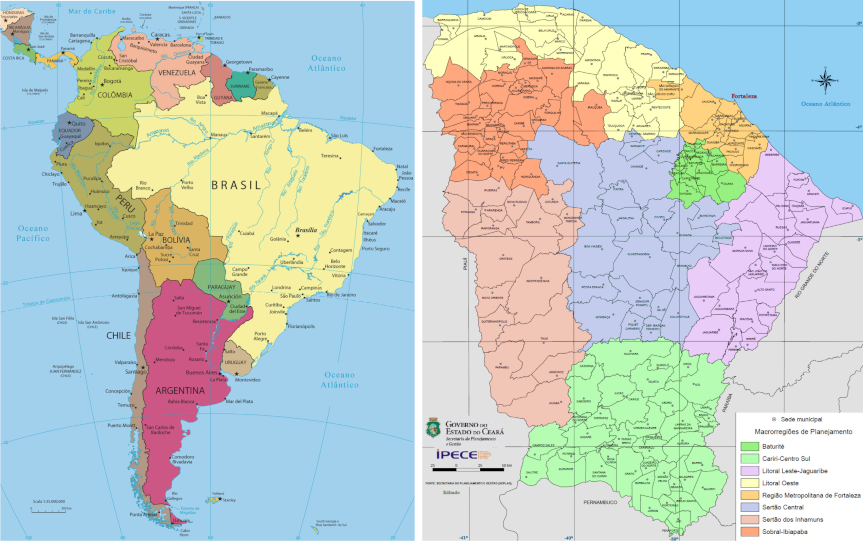}}
\caption{Left: Map of South America, with Brazil and
the city of Fortaleza
visible in the north. Right: Map of Ceará, where
one can spot Sobral, Fortaleza, and Camocim,
the nearest port to Sobral.}
\label{brazil}
\end{figure}

\begin{figure}[h]
{\includegraphics[scale=2.02]
{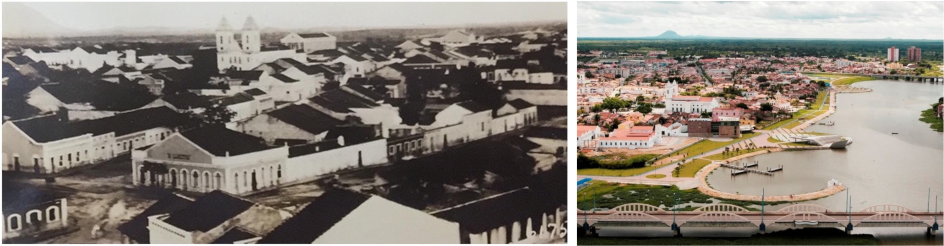}}
\caption{Left: Sobral in 1919 with the main church of Nossa Senhora
da Conceição seen from the church of Patrocínio with a zoom.
The Jockey Club where the English and American tents were
mounted is just beyond the right boundary of the photograph.
Courtesy of Carnegie Institute.
Right: Sobral city today
with the main church of Nossa Senhora da Conceição seen
from near the river Acaraú, with the river itself on the picture.
The old race horse track of the Jockey Club
started from the riverbank at
the constructed circular bay seen in the figure and 
ran perpendicular to the river.
}
\label{NossaSenhoradaConceicao}
\end{figure}

Sobral is located 230 Km west
from Fortaleza, the capital of the state of Ceará,
and the port nearest to it is Camocim,
see Fig.~\ref{brazil}.
Sobral is crossed by the river Acaraú.
In 1919 it was a small town of 10 thousand people,
it is now
a fine city with a population of 200 thousand. 
It is worth to compare  photos of Sobral city in 1919 and in 2019,
see Fig.~\ref{NossaSenhoradaConceicao}.

\begin{figure}[h]
{\includegraphics[scale=1.10]{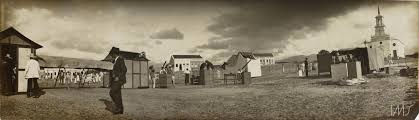}}
\caption{Observatory camp of the Brazilian expedition with the church of
Patrocinio in 1919. Courtesy of Observatorio Nacional, Rio
de Janeiro.}
\label{ObservatorycampBrazilwithPatrocinio}
\end{figure}

\begin{figure}[h]
{\includegraphics[scale=1.45]{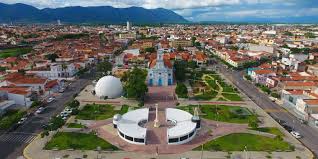}}
\caption{Museum of the Eclipse, the planetarium, and church
of Patrocinio today. The monument in between the two half moons
of the museum was built in 1923 to celebrate
the 150 years of the establishment of Sobral.}
\label{museumplanetariumchurch}
\end{figure}

\begin{figure}[h]

{\includegraphics[scale=1.05]{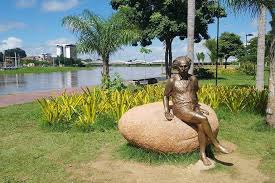}}
\caption{Einstein in Sobral: Einstein's statue inaugurated
in 2019 for the one hundred years of the eclipse.}
\label{EinsteininSobral}
\end{figure}

The Brazilian astronomers, led by Morize, were also
at Sobral to make observations of the Sun's cromosphere
and corona. They mounted their tents with the telescopes
in front of church of Patrocinio,
see Fig.~\ref{ObservatorycampBrazilwithPatrocinio}.
Church of Patrocinio is now famous because of this.
In 1999, Sobral inaugurated the Museum of the Eclipse
precisely in the place where the Brazilian tents were,
as can be ascertained
from Fig.~\ref{ObservatorycampBrazilwithPatrocinio}
and Fig.~\ref{museumplanetariumchurch}.
The two stylish half moons format of the museum
is due to the architect
from Sobral Antenor Coelho.
On one side of the museum there is a small dome
that contains
the Henrique Morize Observatory with its telescope,
and since 2015 a
planetarium was also installed. It is a true astronomy
park. This year of 2019 a sculpture of Einstein was inaugurated
and laid near the river, see Fig.~\ref{EinsteininSobral}.

The two members of the British expedition, Crommelin and
Davidson, were installed
nearby in the house of the Saboya family, together
with the two members of an 
American team of
geophysicists from the Carnegie Institute
that, jointly with
other teams of
geophysicists
spread over several countries within the belt
of the eclipse, wanted to measure the magnetic
field of the Earth and the electricity of the air
during the eclipse.
The Saboya
house is not in front of the church of
Patrocinio, it is a three minute walk from
it, to the back and to the right of 
Fig.~\ref{museumplanetariumchurch}, i.e., to
the south of the church.
In front of the Saboya house
passed a horse race track
that started in the Aracaú's riverbank,
had 700 meters length,
and belonged to the Jockey Club of Sobral.
Since there
were no races in the foreseeable future the
British and American teams
mounted the telescope tents precisely there.
The Jockey Club,
 also called the 
Derby Club, has moved somewhat further to the
northeast of the town along the river Aracaú.

\newpage
\centerline{}

\newpage
\subsection{Principe}

\begin{figure}[h]
{\includegraphics[scale=0.8]{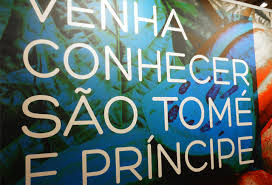}}
\caption{Come and see S\~ao Tom\'e e Pr\'incipe.}
\label{stp3}
\end{figure}

The track of the eclipse would continue
its trajectory through a narrow band on
Earth. After Brazil, it would 
cross the Atlantic, 
would touch Cape Palmas in Liberia,
pass through the island 
of Principe and spread through
mainland Africa until disappearing in the
east. The Joint Permanent Eclipse Committee
with Dyson and Eddington in command
chose
the island of Principe.
Principe,
belongs to the S\~ao Tom\'e e Principe
archipelago,
was a Portuguese territory in 1919
that became independent in 1975.
Why Principe was chosen instead of
other point in mainland Africa is
motif for speculation.
It is known that
Principe was chosen instead of 
Cape Palmas because it was thought
that the chances of having good weather
in Principe were better than in Cape Palmas,
although on the day of the eclipse
Cape Palmas had a sunny day,
as reported by one of the geophysicists
from the Carnegie Institute
that composed the several teams spread
along the eclipse's belt set
to measure the magnetic
field of the Earth \cite{bauer}.
It is also known that
Eddington and Davidson had passed in Lisbon
in 1912 on the way to the total eclipse in
Passa Quatro, Minas Gerais, Brazil, where
they wanted to observe
the solar corona.
In this
passage through Lisbon, Eddington met
Campos Rodrigues and Frederico Oom,
director and vice-director
of the Lisbon Observatory, respectively,
and surely considered himself at ease
to ask for observing the 1919 eclipse
in Principe. Moreover, the eclipse would strike
Gabon, a French colony at the time,
and certainly would be more delicate
to pass by the French sensitivities,
as the Europe powers were in another peak of national
fanaticism. Thus, in the end Principe
was the best choice, see
Fig.~\ref{stp3}. 

\begin{figure}[h]
{\includegraphics[scale=2.00]{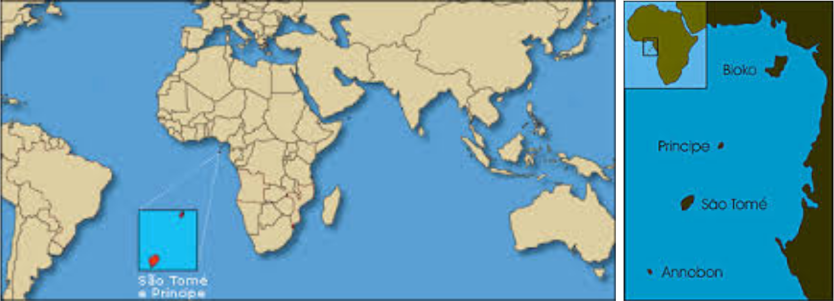}}
\caption{Left: The location of Principe in the Guinea Golf off the
western equatorial coast of Africa. Right: A close up.}
\label{stp2}
\end{figure}

\begin{figure}[h]
{\includegraphics[scale=0.42]{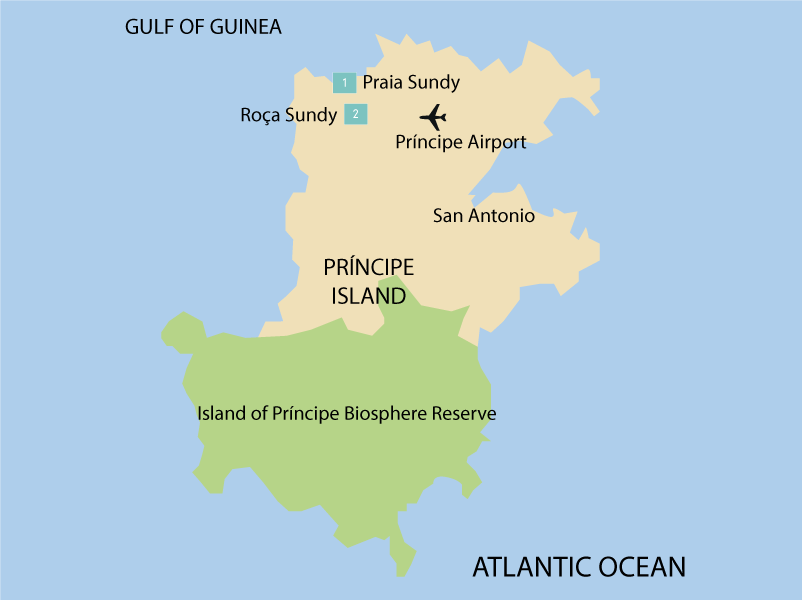}}
\caption{Map of Principe showing the location
of Santo Antonio and
of Ro\c ca Sundy.}
\label{mapofprincipe}
\end{figure}

\begin{figure}[h]
\vskip -0.1cm
{\includegraphics[scale=0.45]{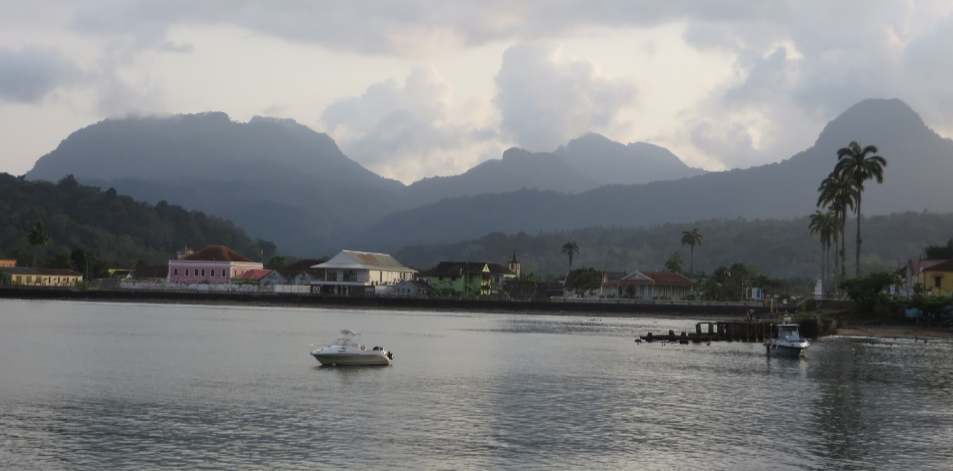}}
\caption{Santo Antonio bay.}
\label{SantoAntoniobay}
\end{figure}

Principe is in the Gulf of
Guinea off the western coast of Africa
just north of the equator, see 
Fig.~\ref{stp2}.
A map of Principe is shown in Fig.~\ref{mapofprincipe}
with a few important locations, namely, Santo
Antonio and Ro\c ca Sundy.
In Santo Antonio, the capital of Principe, see Fig.~\ref{SantoAntoniobay},
Eddington and Cottingham were received
by several personalities of the island.


One of the personalities that took the British
around to see the best site to mount the
telescope was Jerónimo Carneiro, a  landowner with headquarters at
Roça Sundy. 
\begin{figure}[h]
{\includegraphics[scale=2.00]{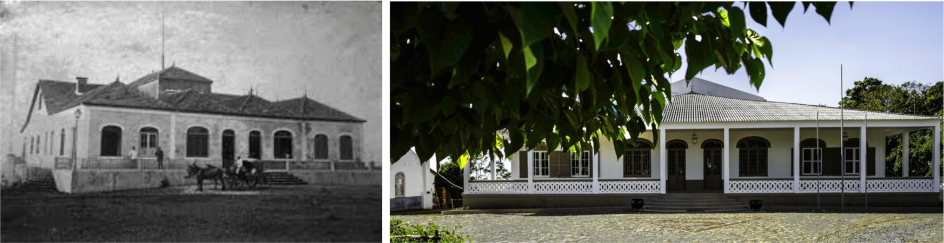}}
\caption{Left: Ro\c ca Sundy house then.
Right: Ro\c ca Sundy house now.}
\label{rocaSundycasa}
\end{figure}
In 1919 he was a young man of
26 years old and had  inherited the land from
his father and from
his grandfather. His grandsons, live in
Lisbon, and remember
him well, who is also nowadays
remembered by the local
old people of Principe. With the independence
in 1975, the land became a state land
and the family Carneiro did not return there.
The land
passed through a river, called the river Sundim, that
was in a region that belonged to
Senhor Dias probably in the beginning
of the 19th century.  Local people rounded his name
for Sundim. So, from Senhor Dias it passed to Sundim,
from Sundim it passed to Sundi and then from Sundi
it changed to
Sundy. It is possible that the first time it
was written as Sundy, clearly an English spelling,
was by Eddington when he was reporting 
on the eclipse preparations and results. 
With all the best
possible infrastructure
that could be offered in
Principe the British opted to stay at Roça Sundy.
\begin{figure}[b!]
{\includegraphics[scale=0.18]{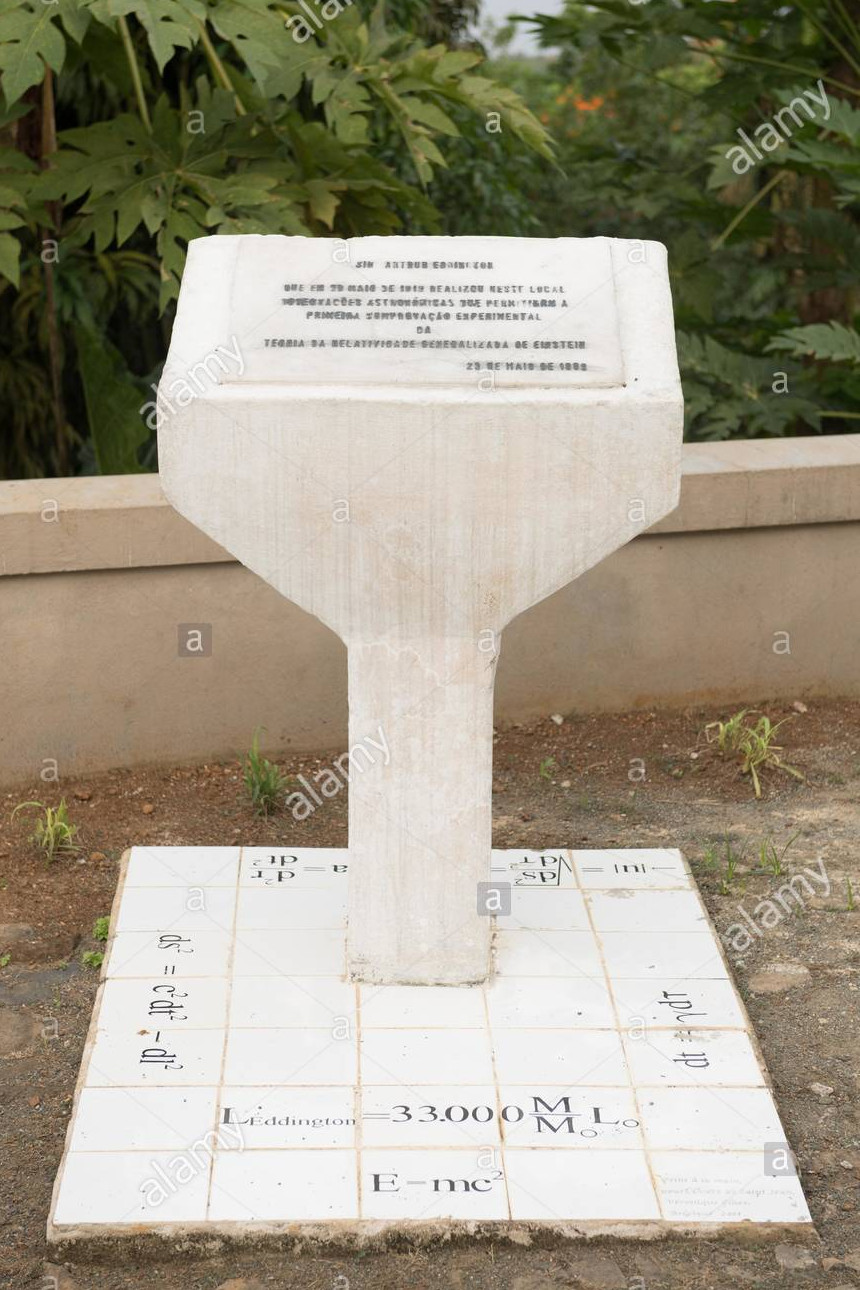}}
\caption{A stone of 1989 in Roça Sundy celebrating Eddington's
confirmation of general relativity.}
\label{sundyinfo4}
\end{figure}
\begin{figure}[h]
{\includegraphics[scale=0.15]
{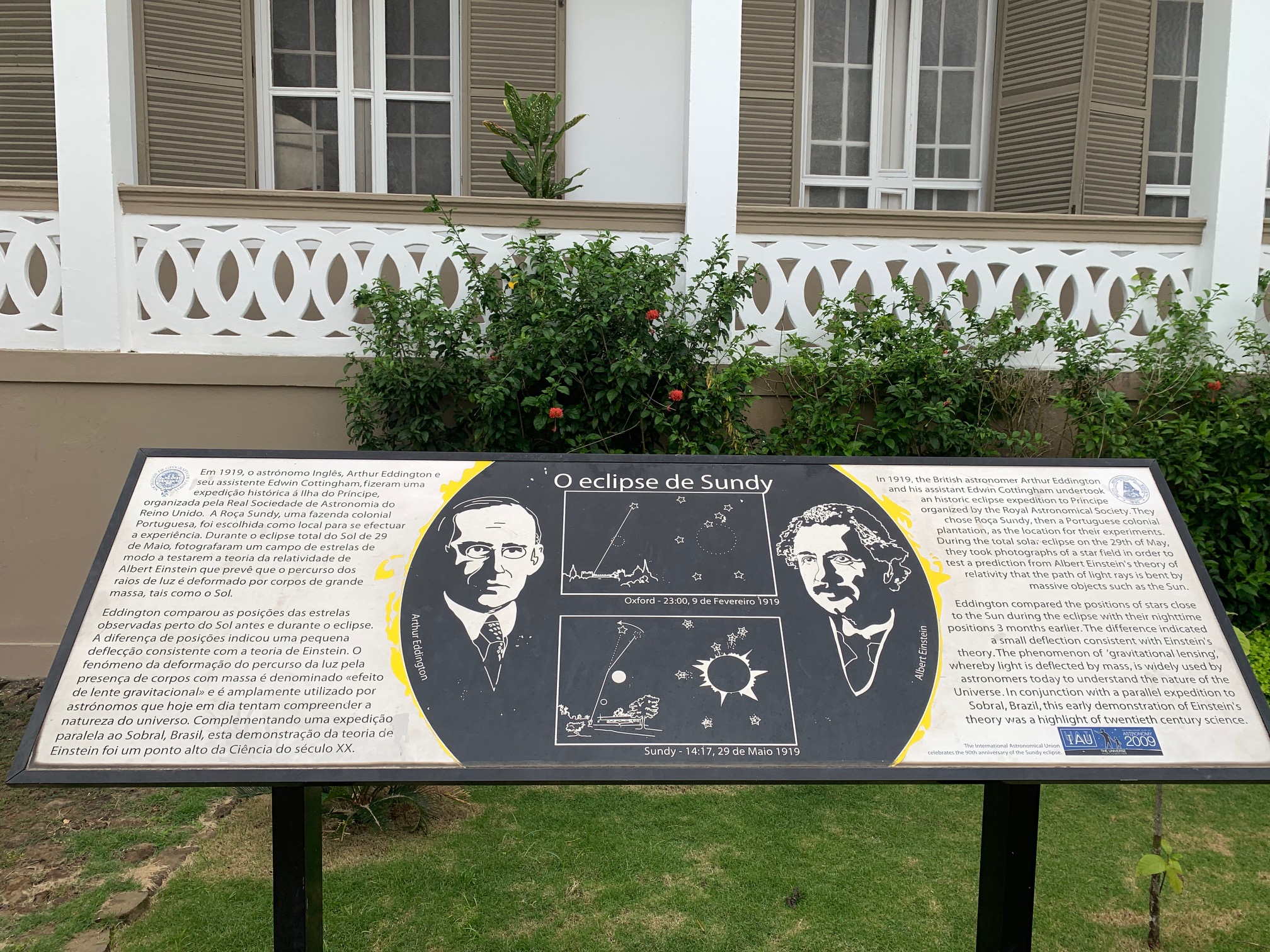}}
\caption{A plate in Roça Sundy containing useful historical and
scientific information about the findings of the 1919 eclipse.}
\label{sundyinfo5}
\end{figure}
In Roça Sundy, in the surroundings of the house
there is a stone commemorative of the
1919 eclipse and Eddington's observations that
confirmed the generalized
theory of relativity,
see Fig.~\ref{sundyinfo4}. On its base there are some
mathematical formulas, like the Eddington 
luminosity expression, a line element
representing curved spacetime, and the gravitational
redshift formula. 
There is also a plate, now in front of one
of the sides of the house, see Fig.~\ref{sundyinfo5},
with a well-written
small overview of the main achievements
of the eclipse and an informative
figure in the middle explaining the
phenomenon of light deflection and
with pictures of Eddington and Einstein.

\section{The theory: what general relativity says about
light deflection}

\begin{figure}[h]
{\includegraphics[scale=0.80]{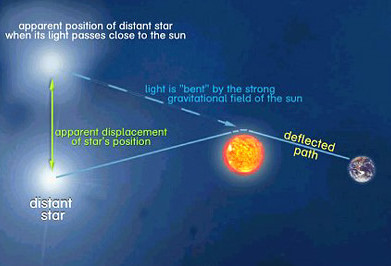}}
\caption{A drawing showing the light deflection
phenomenon. Courtesy of Light Website.}
\label{lightdeflectionallfancy}
\end{figure}

In 1907 Einstein understood that in an accelerating field,
e.g., in an accelerating elevator, the trajectory of a
light ray would be bent, and through the equivalence
principle inferred that light would be as well deflected
in a gravitational field of a star, like the
Sun for instance
\cite{einstein1907}, 
see Fig.~\ref{lightdeflectionallfancy} for a generic sketch
of the phenomenon.
The calculation gave that a light ray would be deflected
by an angle $\delta$ given by the expression
${\delta}=\frac{2GM}{c^2 D}$,
where $G$ is Newton's constant, $M$ is the mass of
the star, $c$ is the velocity of light,
and $D$ is the distance of closest
approach of the light ray to the star
\cite{einstein1911}.
This was a surprise because light
was an electromagnetic wave and it was thought
that waves would have zero mass
and in the context of Newtonian
gravitation would not be influenced by a
gravitational field.
Even with the appearance of
the special relativistic mass formula
$m=E/c^2$, with $E$ being the energy of the wave,
in which case the lightwave
would have precisely the effective mass $m=E/c^2$
and would suffer gravitational attraction, it seems
no one thought of that gravitational
influence on the wave, it was only
through Einstein's  equivalence principle
that the idea and the effect of light deflection
emerged \cite{einstein1907,einstein1911}.

Unbeknown to Einstein,
and to everybody else
interested in light deflection at the time,
this deflection had been calculated
in 1801 by
Soldner \cite{soldner,jaki}, a German astronomer,  using
Newtonian gravitation and considering light
as a particle, yielding the same result
as using the equivalence principle.
Indeed, the
inertial mass term in Newton's 
second law,
cancels with the
the
gravitational mass
in Newton's law of gravitation, this being the essence of
the equivalence principle, and the equation of motion 
holds for any
particle, including particles
with zero mass in the limit. After all gravitation
is universal.
The only  thing 
Soldner's assumed
in the
computation, apart from Newton's laws,
was that the particle travels with the
velocity of light.
The calculation had also been 
done before by Cavendish
but not published \cite{cormmach1968}.
In 1704, Newton 
in the first query
of his book Opticks, questions 
``Do
not Bodies act upon Light at a distance, and by their action bend its
Rays; and is not this action strongest at the least
distance?''
It is 
often said that
Newton 
foresaw gravitational deflection in this  query.
It seems that this interpretation is not correct. 
In the query he never
mentions gravitation, indeed the book is on opticks, 
so it is almost certain that
he was thinking in other interactions
of light with matter.
When seventeen years before, he was writing
the Principia on the laws of
motion and gravitation, he would have thought
that light could be deflected by a gravitational
field, but he may have dismissed it as trivial, 
as gravitation was universal, 
and in his theory light had no special 
category, its motion being a particular case
of unbound hyperbolic trajectories
of particles. 
Clearly, if he had thought it important he
would have done the 
calculation explicitly in the Principia
or even afterwards.

Now, by 1915 there were competing
relativistic theories of gravitation.
Nordstr\"om's scalar theory \cite{nord},
on which Einstein was very interested
although it yielded the wrong value
for Mercury's precession, gave zero
deflection for light travelling  in
a gravitational field,
the same value as in Newtonian gravitation
considering light
as a wave with no coupling to it,
and the entwurf theory of Einstein and Grossmann
predicted, as the equivalence principle
calculation had done, again
the Newtonian deflection value of Soldner
\cite{einsteingrossmann1914,einstein1914aentwurfdeflection,
einstein1914b}.
So, when Einstein had the 
equations of general relativity ready 
he of course thought important to
redo the light deflection 
due to the Sun's gravitational
field and got ${\delta}=\frac{4GM}{c^2 D}$
\cite{einstein1915perihelion}.
This is twice the result he got using
simply the principle of equivalence.
Let us see how one obtains this $\delta$.

\begin{figure}[h]
{\includegraphics[scale=0.55]
{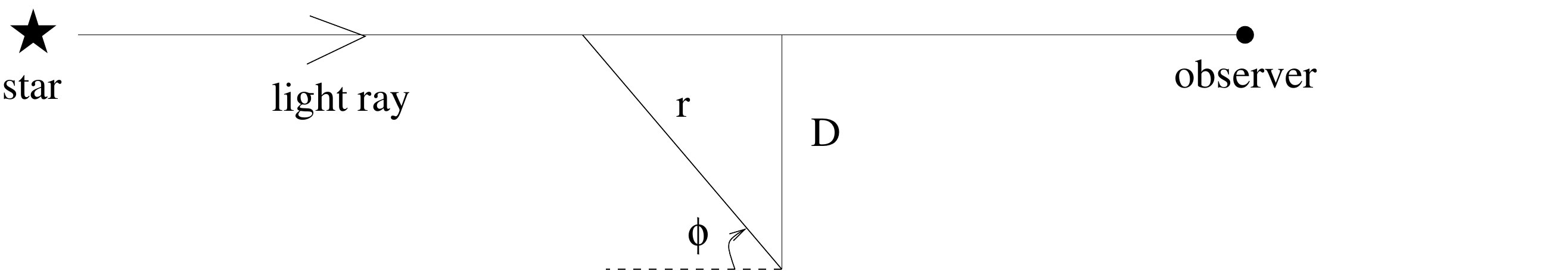}}
\caption{No mass no light deflection.}
\label{lightdeflectionzerowithoutmass}
\end{figure}
If there is no central mass, i.e., no Sun,
then the light ray propagates in a straight line,
see Fig.~\ref{lightdeflectionzerowithoutmass}.
Clearly, in polar coordinates a straight line has
equation $r\sin\phi=D$, where $D$ is the closest approach
distance from the center of coordinates.
It is sometimes better to use the variable $u=\frac1r$
in which case the straight line equation is 
$u=\frac{1}{D}\sin\phi$.

Now, when the center of the polar coordinates has a central mass,
e.g., the Sun, then the problem changes and a light trajectory gets
curved.  The spacetime for a central mass $M$ is given in 
spherical coordinates
$x^a=(t,r,{\theta},{\phi})$, where $t$, $r$, $\theta$, and $\phi$, are
   the time, radial, and angular coordinates, respectively,
   by the Schwarzschild line element
interval
\begin{equation}
ds^2=\left(1-\frac{2M}{r}\right)\,dt^2
-\frac{dr^2}{1-\frac{2M}{r}}
-r^2\left(d\theta^2+\sin^2{\theta}\,d{\phi}^2\right)\,.
\label{metriclightdeflection}
\end{equation}
When one writes $M$ in Eq.~(\ref{metriclightdeflection})
one should bare in mind that one should have written $GM/c^2$,
or conversely, if one uses units in 
which  $G=1$, $c=1$, than one can stick to $M$ simply.
So, $M$ is the geometrical mass with units of length. 
For the Sun, in kilograms $M=2\times10^{30}\,$Kg, and
in kilometers $M=1.5\,$Km.
The light trajectories follow geodesics, paths of
minimal interval.  One can think that there is a parameter $\lambda$
that runs along a geodesic and so the geodesic trajectory can be
written as $x^a=x^a({\lambda})$, i.e.,
$x^a=(t(\lambda),r(\lambda),\theta(\lambda),\phi(\lambda))$.
\begin{figure}[h]
{\includegraphics[scale=0.55]{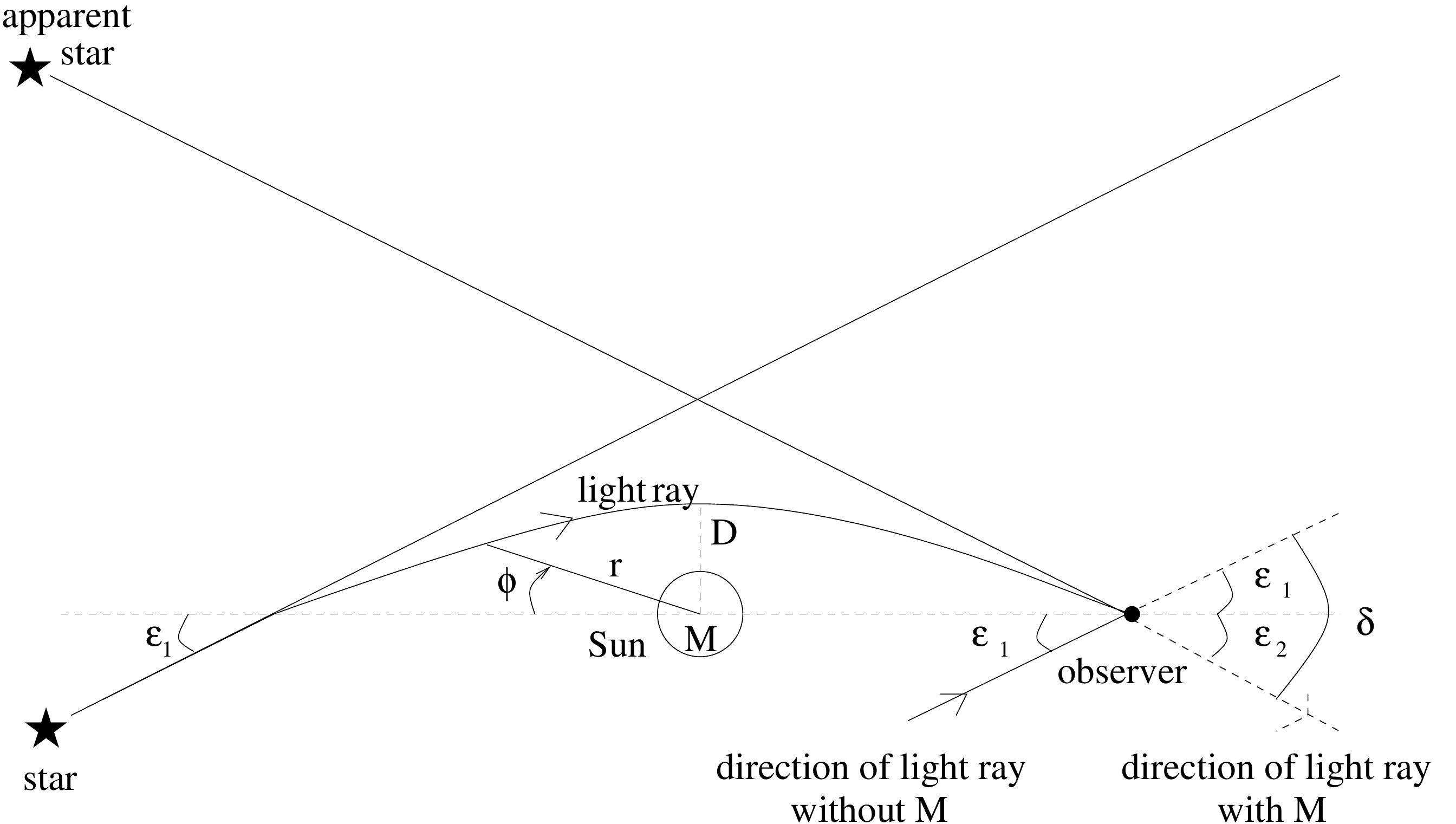}}
\caption{
When there is an object of mass $M$ like the Sun there is light
deflection.
Light rays come from a star at infinity. If there is
no mass $M$ they continue their travel in straight lines.
Since the star is at infinity
all these light rays are parallel and hit the line
Sun-Earth at an angle
$\varepsilon_1$, in particular one of them hits
the observer at that angle. 
In the presence of a mass $M$
these light rays get deflected and
one of them 
hits the observer at an angle $\varepsilon_2$.
The incoming light ray is thus deflected
by an amount $\delta=\varepsilon_1+\varepsilon_2$.
Therefore the  star has its position in the sky
shifted by $\delta$.}
\label{lightdeflection}
\end{figure}
One can 
assume that the light propagation is in the plane
${\theta}=\frac{\pi}{2}$,
so the angle $\theta$ along the 
light ray
is inert, it does not change.  
Two equations for the light 
geodesic can be shown to be 
$(1-\frac{2M}{r})\dot t=k$ and $r^2 {\dot\phi}^2=b$,
where 
a dot means derivative with respect to $\lambda$, and
$k$ and $b$ are constants, actually constants
of integration.
Since the
trajectory is light-like one has $ds^2=0$.
Thus, a third 
equation for the geodesic can be find 
by putting 
$ds^2=0$ in  Eq.~(\ref{metriclightdeflection}),
and dividing each term of it
by $d\lambda^2$, one gets
$(1-\frac{2M}{r})\dot t^2-\frac{\dot r^2} {1-\frac{2M}{r}}-r^2
{\dot\phi}^2=0$,
where $\dot\theta=0$ has been assumed.
So putting the first two equations onto the third 
yields the
geodesic equation $\frac{k^2}{1-\frac{2M}{r}} -\frac{\dot
r^2}{1-\frac{2M}{r}} -\frac{b^2}{r^2}=0$.
Now, define again 
our variable $u$ as $u=\frac{1}{r}$. Then, since 
by definition
$\dot
r=\frac{dr}{d{\lambda}}$, one has
$\dot
r=\frac{dr}{d{\lambda}}= \frac{d(1/u)}{d{\lambda}}=-
\frac{1}{u^2}\frac{du}{d{\phi}}\frac{d{\phi}}{d{\lambda}}
=-b\frac{du}{d{\phi}}$, i.e.,
one trades a derivative with respect to
$\lambda$ by a derivative with respect to
$\phi$.
Then the geodesic equation above turns into
$\left(\frac{du}{d\phi}\right)^2+u^2=2Mu^3+ \frac{k^2}{b^2}$. Usually
one wants to integrate, but here we take the derivative in relation to
$\phi$ to get a better equation, which gives
$\frac{d^2u}{d\phi^2}+u=3Mu^2$, as  
$\frac{du}{d\phi}$ appears in all terms and can been eliminated.
No central mass, $M=0$, and the equation is
$\frac{d^2u}{d\phi^2}+u=0$, the solution being
obvious, it is a sine, i.e., $u=\frac{1}{D}\sin\phi$. 
Or putting back $u=\frac1r$ get
a straight line in
polar coordinates $r\sin\phi=D$, as we have shown above.
Now $\frac{M}{D}$ is small, as $M$ is the geometric
mass of the Sun, a small quantity, and 
$D$ is the distance of closest approach
of the light ray to the Sun's surface, a much bigger quantity.
For $\frac{M}{D}$ small one uses perturbation methods and finds
the solution
$u=\frac{1}{D}\sin\phi +\frac{M}{2D^2}\left(3+\cos2\phi\right)$.
The light ray comes from the star,
see Fig.~\ref{lightdeflection}, at 
$r=\infty$, i.e., $u=0$, and with
$\phi=-\varepsilon_1$ say,  so we 
get from the solution $0=-\varepsilon_1+4\frac{M}{2D}$,
i.e., $\varepsilon_1=\frac{2M}{D}$.  
The light ray goes to the Earth,
see Fig.~\ref{lightdeflection}, assumed at $r=\infty$,
so $u=0$,  and with $\phi=\pi+\varepsilon_2$ say,  so get
from the solution
$0=-\varepsilon_2+4\frac{M}{2D}$, i.e., $\varepsilon_2=\frac{2M}{D}$.
We used the approximation $\sin\varepsilon=\varepsilon$ for
$\varepsilon\ll1$, as is the case here.
Now, the total angle of deviation is
$\delta=\varepsilon_1+\varepsilon_2$,
see Fig.~\ref{lightdeflection},  so
summing the two results above one gets $\delta=\frac{4M}{D}$.
Restoring
$G$ and $c$ one has that the formula for the deflection taken from
general relativity is
\begin{equation}
{\delta}=\frac{4GM}{c^2 D}\,.
\label{hyperboliclaw}
\end{equation}
This is the $1/D$ law for light deflection, 
i.e., an hyperbolic law.
Let us put numbers. The geometrical mass of the Sun is
$M=1.5\,{\rm Km}$, and
for a light ray that grazes
the Sun's surface, $D$ is the Sun's radius,
i.e., $D=7\times10^5\,{\rm Km}$. So $\delta=0.85\times10^{-5}\,{\rm
rad}= 1.75\,$ arcseconds.  The general relativistic deviation for a
light ray that passes tangent to the Sun's surface is 1.75
arcseconds and so an
apparent star shifted
by this amount is seen in the sky
instead of the star in its true position.
Newtonian gravitation predicts half of that value.  To
see this one can follow Soldner's calculation, or for that matter
Einstein's calculation using the equivalence principle, but simpler
here is to consider that Newtonian gravitation can be mimicked by a
line
element interval of the form $ds^2=-\left(1-\frac{2M}{r}\right)\,dt^2
+{dr^2} +r^2\left(d\theta^2+\sin^2{\theta}\,d{\phi}^2\right)$, where
the term in front of $dt^2$ takes care of the Newtonian gravitational
potential $\frac{M}{r}$, and there is no term
$\frac{1}{1-\frac{2M}{r}}$ as space is flat. Then redoing the
calculations above, instead of having 
${1-\frac{2M}{r}}$ appearing twice as in 
Eq.~(\ref{metriclightdeflection}), one has ${1-\frac{2M}{r}}$
appearing only once, yielding
half the value, ${\delta}=\frac{2GM}{c^2 D}$, so that
for a light ray passing at the Sun's rim ${\delta}=0.875\,$ arcseconds.
Light deflection is treated
in most textbooks 
on general relativity.

\begin{figure}[h]
{\includegraphics[scale=0.5]{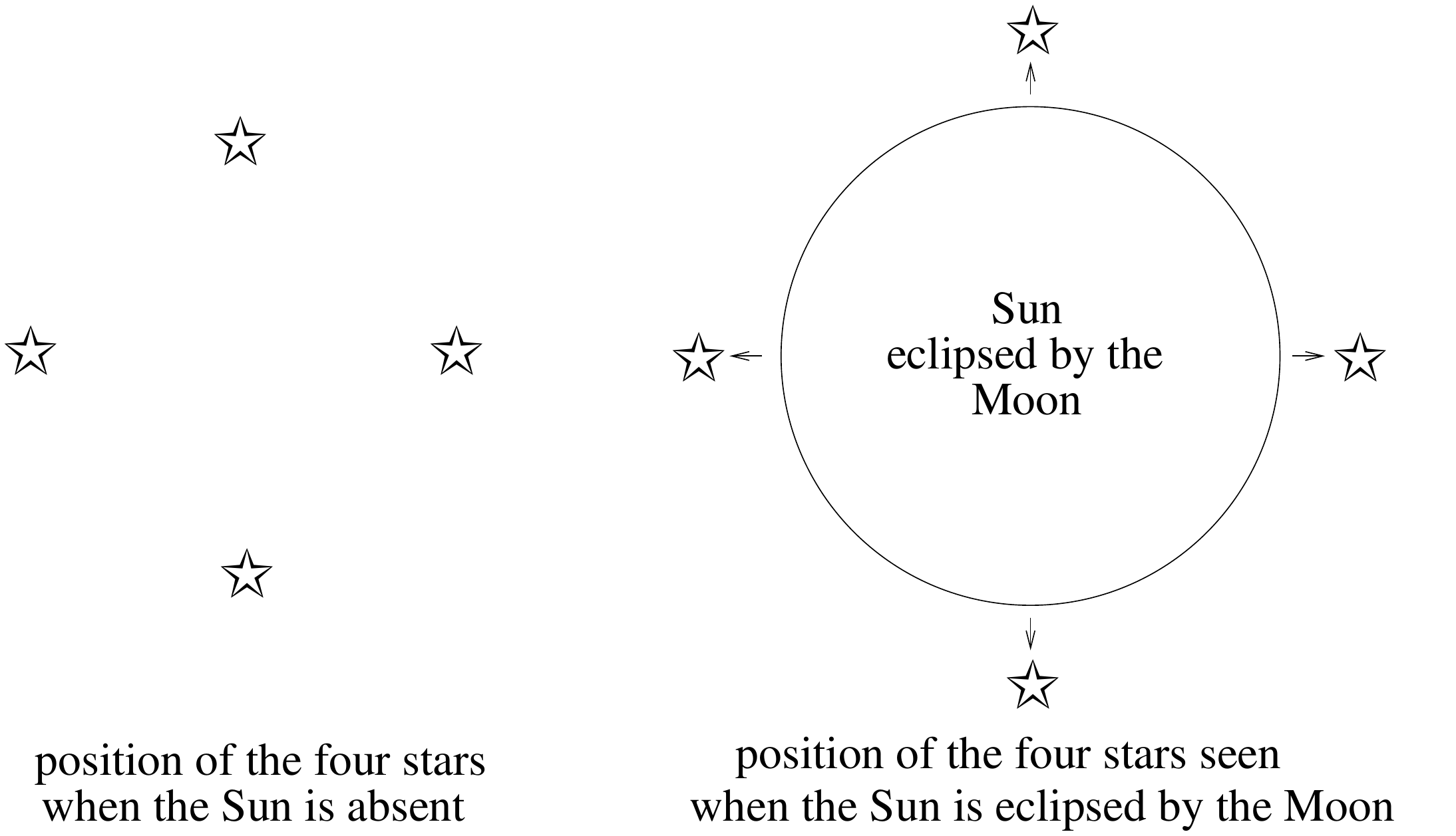}}
\caption{Left: Field of four stars. Right:
The Sun is eclipsed by the Moon and the
field of four stars pops out
in accord with the $1/D$ law for gravitational
light deflection.}
\label{fieldoffourstarsbehindthesun}
\end{figure}

Admit that the Sun is eclipsed by the Moon, as in
Fig.~\ref{fieldoffourstarsbehindthesun}, and
that there are four stars just at the rim of the Sun. Then due
to the gravitational deflection effect the four field stars will pop out
just outside the Sun's rim.
This effect always exists, with or without
eclipse, but only with an eclipse can the 
stars, and so the effect, be
observed. Measure the positions of the 
four stars when the Sun is not there
and measure the positions of the four stars 
when the Sun is there, subtract the results 
for each star, and obtain $\delta$ for each star. 
Compare this observed $\delta$ with the
$\delta$ predicted by
the theories of gravitation, namely,
the zero value 
of
Newtonian gravitation
with no coupling to lightwaves
or Nordström relativistic scalar theory
of gravitation, the half value 
of Newtonian gravitation
and treating
light as a particle
as considered
by Soldner
or considering
light as a wave with an effective mass
$E/c^2$, and
the full value of general
relativity. 
This concept is what is needed to understand 
the light deflection test
of theories of gravitation,
in particular of general relativity,
in eclipse observations. So now we turn into 
the heart of the 1919 eclipse and 
its light deflection.

\section{The 1919 eclipse: Preparations, the day, data analysis,
results, and history}

\subsection{Preparations}

Having understood that a light ray from a star behind the Sun is
deflected by its gravitational field \cite{einstein1907,einstein1911},
Einstein immediately started to look for astronomers that could measure
the effect.
Freundlich, a young German astronomer,
got very interested, and tried first to obtain 
from other astronomers around the world pictures of
past eclipses, but these were not of the quality
required. 
It was necessary to wait for a total eclipse of
the Sun.
The American astronomer Charles Perrine,
originally from Lick Observatory and now director of
the Cordoba Observatory in Argentina, heard from
Freundlich the interest and importance of testing
the light deflection in the gravitational field of the Sun.
In October 10, 1912, there was an eclipse that passed in
Minas Gerais, Brazil,
and Perrine went there to observe the predicted Einsteinian
equivalence principle half effect
\cite{perrine1923}.
Eddington and Davidson happened also to be there
but to observe the solar corona. At this time,
Eddington was not familiar with Einstein's ideas.
Many other teams from everywhere were also in
Minas Gerais for the eclipse.  
It is speculated \cite{stachel,warwick} that Eddington heard for
the first time from Perrine the possibility of light
deflection. Eddington was at Passa Quatro and
Perrine at Cristina, both localities in the state of
Minas Gerais separated by 80 Km through the road. It
is indeed possible that they met somewhere in some
city of Brazil before the eclipse, although that is
not registered in the official report \cite{eddington1912}.
Unfortunately, Perrine 
``suffered a total eclipse instead of observing one'',
as he wrote  
after heavy rain set in \cite{perrine1923},
and no light deflection could be checked.
In August 21, 1914, there was a solar eclipse that passed
in Crimea, Russia.
It would test the zero prediction value
in the  case 
that light rays
did not couple to gravitation at all as in
Newtonian gravitation with no
coupling to lightwaves or
the relativistic 
Nordström's theory \cite{nord},
or it would give
the Newtonian value also predicted by the equivalence principle
\cite{einstein1911}
and the entwurf theory
\cite{einsteingrossmann1914,einstein1914aentwurfdeflection},
as Einstein was aware \cite{einstein1914b}.
Or, perhaps, other value.
To test it 
two expeditions were organized
but were a total failure. Freundlich
was made prisoner when he arrived in Russia as the first
world war had just broken out, and Campbell of Lick Observatory
in California, got rained washed on the day.

The eclipses in Brazil and Russia were set
to test the zero value and the half value
prediction, this latter given by the equivalence principle
and the entwurf theory and also 
called  the Newtonian value.
From the end of 1915 onwards, the full value
prediction of general relativity
was also to be tested.
The eclipse of June 8, 1918, that crossed
the USA, could thus serve to test
the zero value,
the half value, and the full value.
The 1918 eclipse passed almost through  
Lick Observatory, but Campbell and Curtis could
not make proper measurements as the
good instruments that went to Russia
were still stuck somewhere because of the first
world war.
The next eclipse would be the one of May 29, 1919.

At the
time, the observation of the Sun's rim through eclipses 
was in its
highest point, and British astronomers and scientists, understanding
the importance of the eclipses,
formed eclipse committees that culminated in
1892 in the Joint Permanent Eclipse Committee
in order to gather a coherent base
of knowledge on eclipses.  Eclipse expeditions
were time consuming and laborious to prepare but to the British
Empire this was no problem, it could put anyone
anywhere.
Dyson, being an  
authority on
eclipses and Astronomer Royal, was a member of the committee.

During the first world war there were no communications between
England and Germany. In England, most scientists,
some 
of the caliber of Lodge, Larmor, and Jeans,
were
vaguely aware of special relativity and even less of general
relativity, they were interested in what was called the electrical
theory of matter
that still included the ether, and 
that perhaps would explain the phenomenon of
gravitation \cite{warwick}.
There were
two exceptions however,
Cunningham from  Cambridge
that taught and published on special relativity, 
and the Lindemanns, father and son,
the first a wealthy engineer and amateur astronomer
of German origin,
the second yet to become a prominent physicist
at Oxford and adviser to
Churchill in the second world war,
that
got interested in the light deflection effect
of general relativity and published a paper
on how one might proceed to measure it from 
daylight photography of stars \cite{lindemanns}.
At about this time, in 1916, 
 Eddington had access to general relativity
 through his friend de Sitter, a Dutch astronomer of Leiden,
colleague of Lorentz and Ehrenfest that had a genuine keen interest in
Einstein's ideas on gravitation.
Earlier, de Sitter had attacked Mercury's perihelion
problem by mixing Newtonian gravitation with special relativity,
and by now he had a good command on general relativity.
Eddington asked de Sitter to write a review on general relativity
to the Monthly Notices of the Royal Astronomical Society
readers which he did in a series of three remarkable
papers. Eddington then immediately understood
the power of the general theory and 
mastered it as quickly as he could, in about two years. He became
particularly interested in the 1919 eclipse test of general
relativity and his first book on general relativity
was published \cite{eddington1918}.

\begin{figure}[h]
{\includegraphics[scale=2.00]{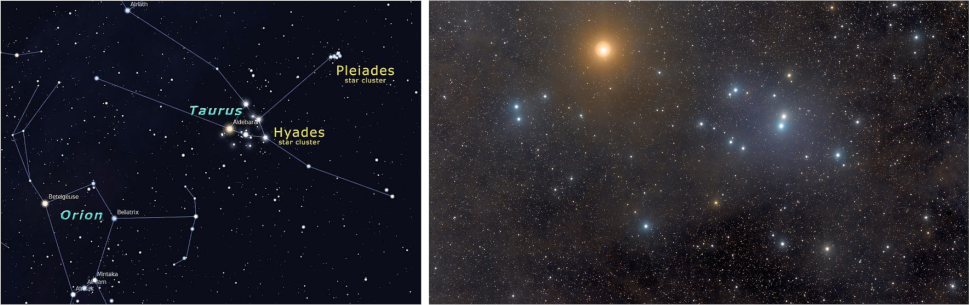}}
\caption{Left: The Hyades open cluster in the Constellation of
Taurus. Right: The Hyades in plenitude.}
\label{hyades2}
\end{figure}

Eddington and Dyson were colleagues that met
several times a month in the Royal Astronomical
Society, i.e.,
in Burlington House, London, 
and in other places around England.
So, briefed by Eddington, Dyson understood the
importance of the 1919 eclipse test and accordingly,
in March 1917, he published
in the
Monthly Notices of the Royal Astronomical Society a paper
with the title
`On the opportunity afforded by the Eclipse of 1919 May 29 of
verifying Einstein's theory of gravitation'' \cite{dyson1919},
in which he highlighted the fact that that eclipse was in
front of the open cluster Hyades, a region of the
sky containing many bright stars, see Fig.~\ref{hyades2}.
A greater number of
bright  stars around the eclipsed Sun
surely improves the probability of obtaining relevant results
for light deflection.

The May 29, 1919, eclipse was thus a great opportunity to test general
relativity. Dyson, in turn, called Eddington to do the observations
for obvious reasons, he was an excellent astronomer and
astrophysicist, and a deep connoisseur of general relativity.  In
addition, Eddington was a quaker, and that religious confession
professes peace and nonviolence. Thus, Eddington, a conscientious
objector, asked to be dispensed of military service, and only by the
intervention of Dyson the exemption from the authorities was conceded
without further ado, with the justification of Eddington's importance
in the expedition.  The preparations started in 1917.  The end of the
first world war in November 1918 enabled definitely the realization of
the new test to Einstein's law of gravitation.  For narratives of this
eclipse and its results see also \cite{crommelin1919naturebefore,
eddington1919,crommelin1919observatory,jjthomson,report1920,eddstg1920,
whitehead,kluber,chandrasekhar,moyer,earman,will2,kennefick}.

\begin{figure}[h]
{\includegraphics[scale=0.75]{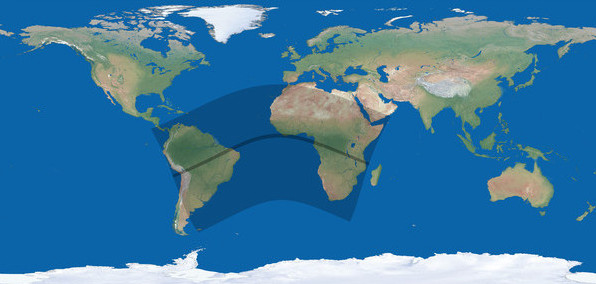}}
\caption{World map with the track of the  May 29, 1919, eclipse.
The black belt 200Km wide yields the places of 
total eclipse. The gray belt is for  partial eclipse.}
\label{worldmapeclipse29may1919}
\end{figure}

The track of the eclipse was known, see
Fig.~\ref{worldmapeclipse29may1919}.
The eclipse would run from west to east, would start at dawn
in west of South American in  Peru,
then traverse Bolivia, 
pass through the north and northeast of Brazil,
cross the Atlantic, hit Cape Palmas in Africa,
reach Principe island, and finally arrive in Africa mainland
until disappearance at twilight in the east in the border
of Tanzania and Mozambique, running across 12 thousand
Km.

Two sites were chosen by the Joint Permanent Eclipse Committee
and accordingly two expeditions were planned.
To have two
expeditions would increase the chances of at least one
being successful.
One expedition would proceed with the astronomers 
Cortie and Davidson
to Sobral, Brazil. Dyson could not go as he had
several other important commitments as Astronomer Royal.
In the last hour Cortie could not depart
and was traded for Crommelin. The Sobral team 
was
pure Royal Greenwich, with Dyson having the
responsibility of supervising
the data analysis and results
after the expedition returned to England.
The other expedition  would proceed 
with
Eddington and
Cottingham to Principe.
The two sites were certainly carefully chosen.
The Sobral site was chosen with the help
of Morize, Eddington and Morize met during
the 1912 eclipse in Minas Gerais.
The Principe site
was chosen for Eddington's acquaintance with
the main Portuguese astronomers, and
for
being the most
convenient at the time taking into
account all factors. 
Why Eddington went to Principe rather than Sobral
is a question that seems beyond
all conjectures.
At last, all was set, and 
Crommelin, Davidson, Eddington, and Cottingham
sailed  on March 8 on the Anselm ship,
passed
in Lisbon
where they were received by Frederico Oom,
went for a city tour \cite{crommelin1919observatory},
and followed to Funchal.

Crommelin and Davidson boarded then for Belém do Pará, stretched up to
Manaus for sightseeing, and went to their destination in the steamboat
Fortaleza, arriving in Camocim, the nearest port, on April 29, and in
Sobral on April 30.  The Sobral expedition had all the support from
the Brazilian authorities.  A team from Observatorio Nacional of Rio
de Janeiro with Morize was present in Sobral.  They accompanied the
stay of the British and mounted its own camping observatory to observe
the solar corona. The Americans from Carnegie had also arrived at
about the same time.
Crommelin describing the arrival
writes \cite{crommelin1919observatory}:
``Several deputations were at the station to welcome us; it must
be confessed that they were expecting Father Cortie,
whose letter expressing his inability to go had never
reached Sobral. However, the welcome
was freely transferred to us.''
This welcome from the Brazilians he restated in
\cite{jjthomson}.
Davidson in the 1920 report published in the
Philosophical Transactions describing the preparations
writes \cite{report1920}: ``We were met at Sobral station by
representatives of both the Civil and Ecclesiastical Authorities,
headed respectively by Dr.~Jacome d'Oliveira, the Prefect, and
Mgr.~Ferreira, and conducted to the house which had been placed at our
disposal by the kindness of its owner, Col.~Vicente Saboya, the Deputy
for Sobral...  A convenient site for the eclipse station offered
itself just in front of the house; this was the race-course of the
Jockey Club, and was provided with a covered grand stand, which we
found most convenient for unpacking and storage and in the preparatory
work.'' 
They
were thus received warmly and treated like princes.

In Funchal,
Eddington
and Cottingham boarded on April 9 in the ship Portugal
to Principe arriving there on April 23.
The Principe expedition had the support
from the  Observatorio
Astronomico de Lisboa 
and from the Portuguese authorities.
There was exchange of correspondence between
Eddington and
Campos Rodrigues and Frederico Oom, to finalize
the logistics.
Eddington, in the report published in 1920 
describing the preparations and the arrival
in Principe, 
writes \cite{report1920}:
``Vice-Admiral Campos Rodrigues and Dr. F. Oom of the National
Observatory, Lisbon, had kindly given us introductions, and everything
possible was done by those on the island for the success of the work
and the comfort of the observers. We were met on board by the Acting
Administrator Sr. Vasconcelos, Sr. Carneiro, President of the
Association of Planters, and Sr. Grageira, representing the Sociedade
d'Agricultura Colonial, who made all necessary arrangements.
The Portuguese Government
dispensed with any customs examination of the baggage.''
Further on he writes:
``We were advised that the prospects of clear sky at the end of May
were not very good, but that the best chance was on the north and west
of the island. After inspecting two other sites on the property of the
Sociedade d'Agricultura Colonial, we fixed on Roça Sundy, the
headquarters of Sr.~Carneiro's chief plantation. We were
Sr.~Carneiro's guests during our whole visit, and used freely his
ample resources of labour and material at Sundy. We learnt later that
he had postponed a visit to Europe in order to entertain us. We were
also greatly indebted to his manager at Sundy, Sr.~Atalaya, with whom
we lived for five weeks; his help and attention were invaluable.''
At Principe, an island with few people, they were
received with great sympathy 
and treated as celebrities and science stars that they were.

The eclipse day was arriving and there are some nice stories
to tell.

\subsection{The day}

In Sobral there are several photographs taken by the Brazilian
astronomers,
the two American geophysicists
from Carnegie, and the two Britons. One of the photographs shows
the three teams, Brazilian, American,
\begin{figure}[h]
{\includegraphics[scale=0.232]{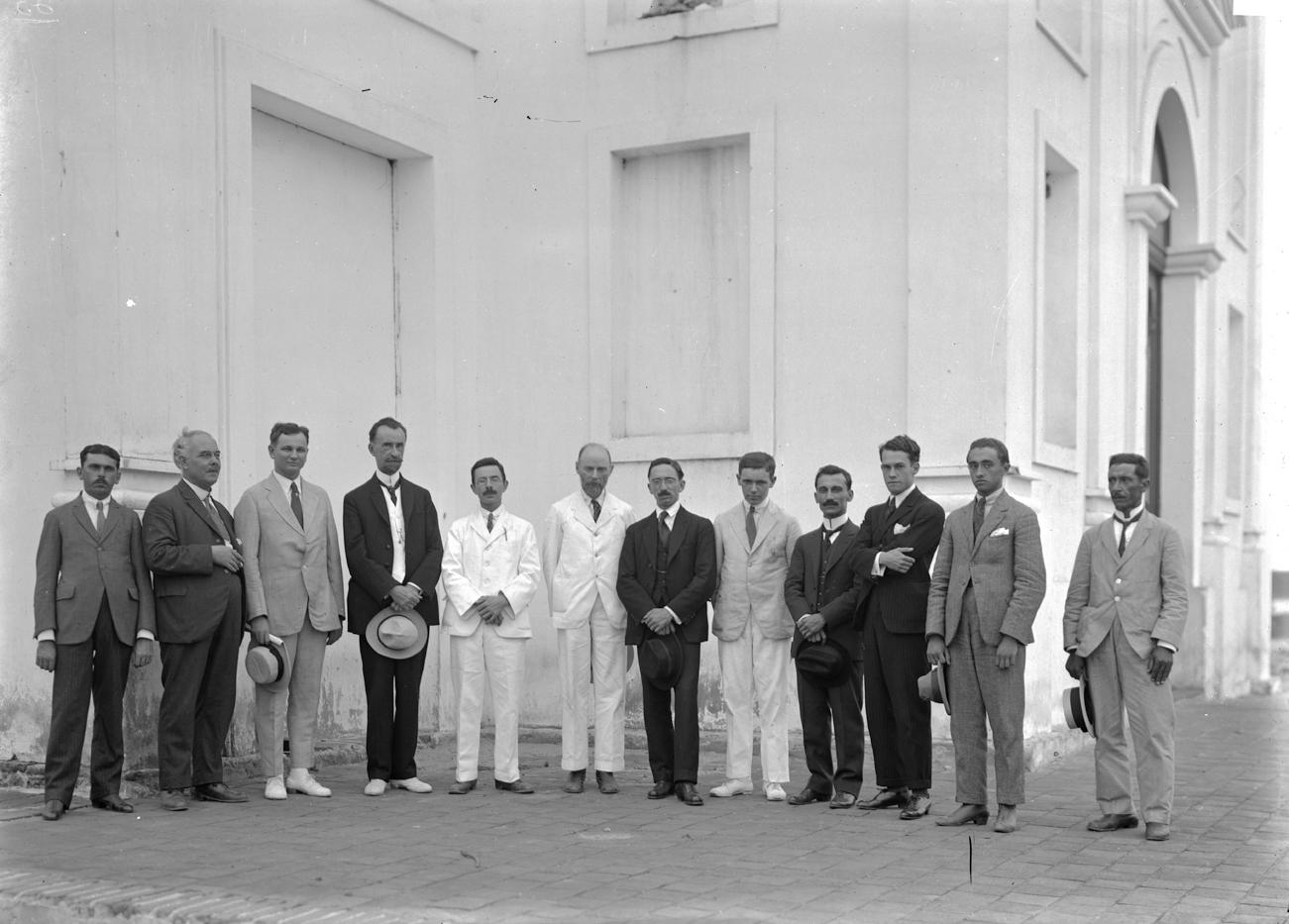}}
\caption{The three eclipse teams, Brazilian, British, and American,
at Igreja do Patrocinio, Sobral. From left to
right: Luiz Rodrigues, Theophilo Lee, Daniel Wise from Carnegie,
Henrique Morize, Charles Davidson, Andrew Crommelin, Allyrio de Mattos,
Andrew Thomson from Carnegie, Domingos da Costa, Lélio Gama, Antônio
Lima, and Primo Flores. Courtesy from Observatorio
Nacional, Rio de Janeiro.}
\label{allatsobral}
\end{figure}
and British, posing together,  see Fig.~\ref{allatsobral}.

\begin{figure}[h]
{\includegraphics[scale=0.11]{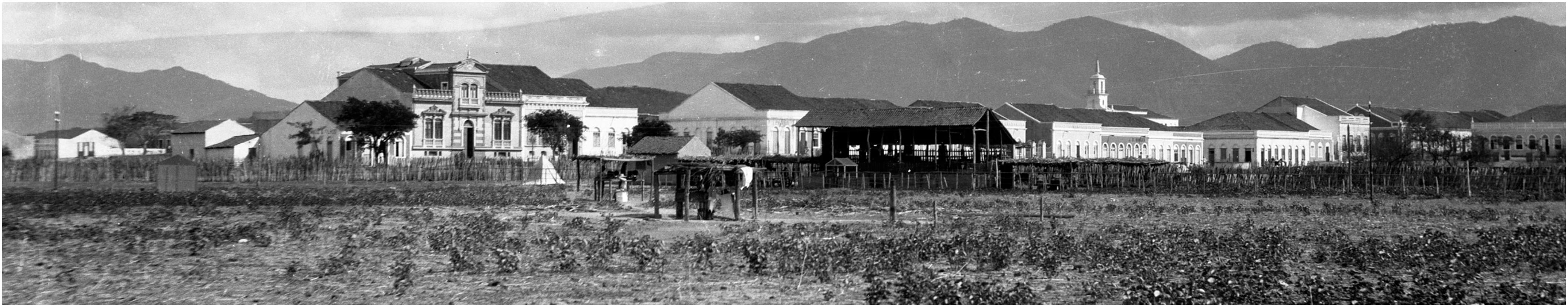}}
\caption{The Jockey race course in Sobral with the tents and the
instruments of the Carnegie team and the British tent. The British
tent is closed and can be seen as a small square on the left
corner. The
Saboya house where the British and Americans stayed is the pleasant
nice looking spacious loft house on the middle left.
Courtesy of the Carnegie
Institute.}
\label{instrumentsatthejockey}
\end{figure}

\begin{figure}[h]
{\includegraphics[scale=2.00]{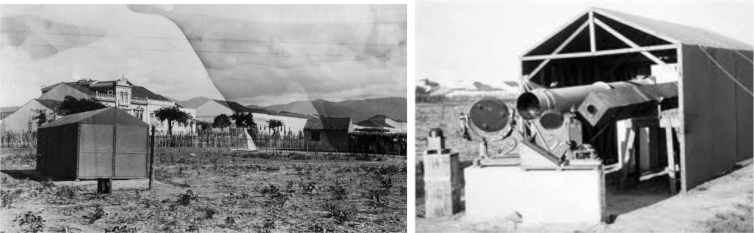}}
\caption{Left: The British tent closed. 
The Saboya house is in the background. Compare with the 
previous photograph.
Right: The British tent opened with the two telescopes used by
Crommelin and Davidson, the 4 inch telescope on the right and the 13
inch astrographic telescope on the left, together with the
respective ceolostats.
The telescopes are pointing eastward as the eclipse
was to be at 9:00am. Courtesy of the Royal Greenwich Observatory.}
\label{telescopesbritish}
\end{figure}

In the Jockey race course, in front of the Saboya house
in which Crommelin and Davidson stayed,
it was mounted
the two telescopes that would serve to make the
light deflection observations.
There was a 4 inch telescope and a 13 inch astrographic
telescope, see Fig.~\ref{instrumentsatthejockey}
and Fig.~\ref{telescopesbritish}.
Since to
mount anew a heavy telescope with all its accessories is a difficult
task, each telescope was coupled to a coelostat, a device commonly
used in eclipses to maintain the telescope rigid.  It consists of a
mirror that reflects the sky field to the telescope itself and turns
slowly with the sky so that the image on the telescope is static.  The
coelostat was conceived by Lippmann in 1895, a French physicist that
got the Nobel Prize in 1908 for inventing color photography.


\begin{figure}[h]
{\includegraphics[scale=0.45]{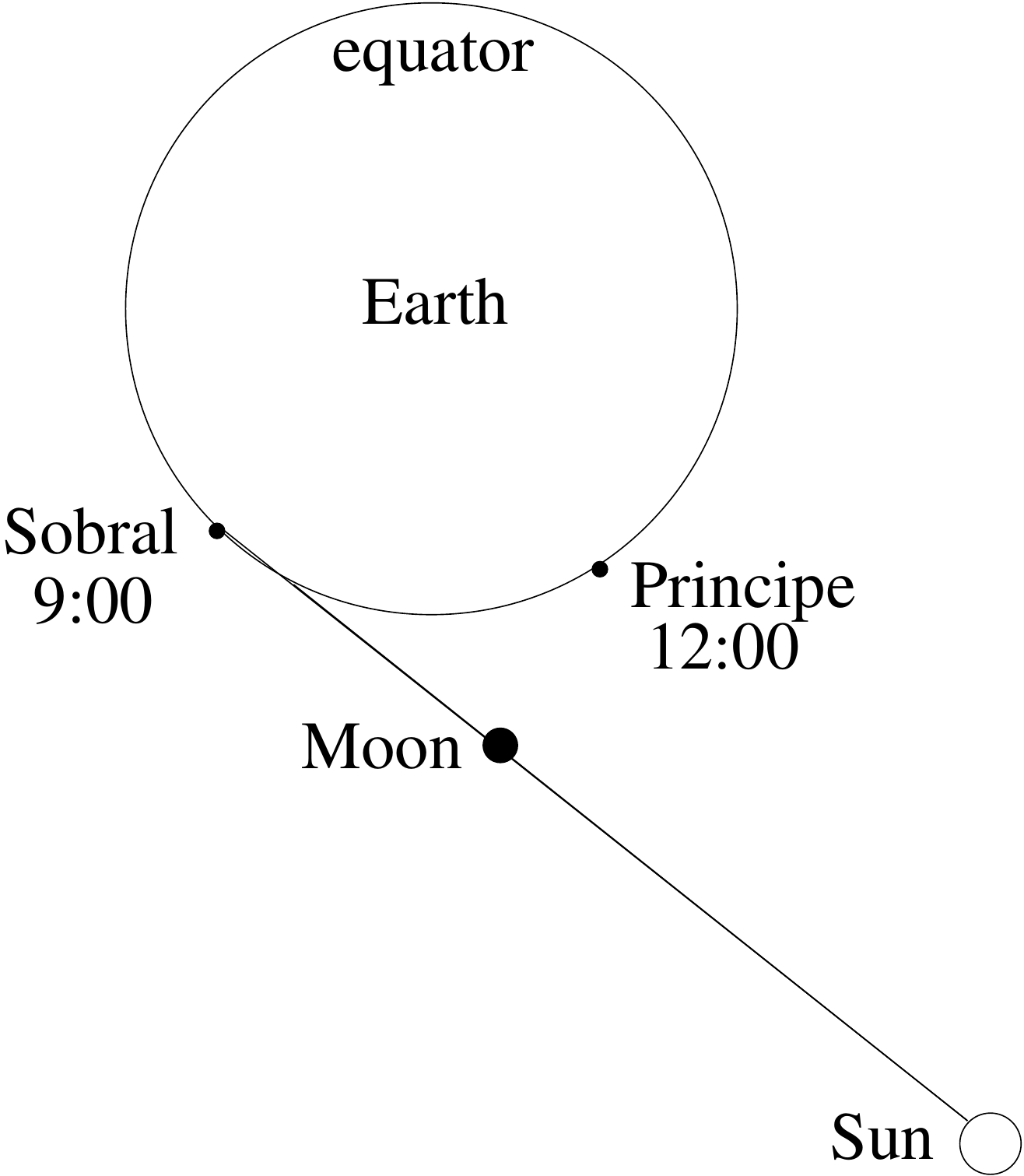}}
\caption{A view from the top of the Earth, Moon,
Sun system yielding an eclipse at 9:00am local time
in Sobral. 
}
\label{eclipsesobral}
\end{figure}

The eclipse in Sobral was in the morning at 9:00am,
see Fig.~\ref{eclipsesobral} for a representation.
Davidson in the report writes
\cite{report1920}: ``As totality approached, the
proportion of cloud diminished, and a large clear space reached the
sun about one minute before second contact.'' Then further on he says,
"The region round the Sun was free from cloud, except for an interval
of about a minute near the middle of totality when it was veiled by
thin cloud, which prevented the photography of stars, though the inner
corona remained visible to the eye and the plates exposed at this time
show it and the large prominence excellently defined.''  And then
commenting on the after the eclipse, 
``On June 7, having completed the development, we left Sobral
for Fortaleza, returning on July 9 for the purpose of securing
comparison plates of the eclipse field.''  So,
Crommelin and Davidson developed the plates, were happy to leave for
Fortaleza, capital of Ceará, to enjoy life on the beaches, and went
back to Sobral to get comparison plates by photographing, the eclipse
field that 7 weeks after the eclipse
could be seen at night before dawn, as the
Earth's translation, or equivalently the Sun's annual movement
along the ecliptic,
modifies slightly but surely the sky, night after night.

Indeed, 
the Sun moves about 1 degree per day on the ecliptic, so that 
365 days after it is back in the same position.
The stars in the celestial sphere stay fixed in the same place.
Now, 24 hours is equivalent to 360 degrees, 
which means that
1 degree per 24 hours is equivalent to (24/360) =(1/15) hours, 
i.e., 4 minutes.
Since this small 1 degree movement of the Sun is in
the same sense of the Earth's rotation, it means
that each day the Sun will be delayed 4 minutes 
per day
relative to the fixed stars, 
for instance the stars of the eclipse. 
Now, the eclipse in Sobral was at 9:00am, 
thus in order that 
the Sun is sufficiently delayed to rise at 6:00am
and the eclipse stars are in the same position
and can now be seen just before dawn, 
one has to have a lag of 3 hours, i.e., 180 minutes, 
which is achieved in 
(180 minutes)/(4 minutes/day)\,=\,45 days, i.e., about 
6 and a half weeks. This is consistent with the 
7 weeks that the British astronomers took to 
return Sobral, where they arrived on
July 9.

Crommelin and Davidson left Sobral on July 22 and arrived back in
Greenwich on August 25.  A report on Sobral is also given in
\cite{eddstg1920}.  For details of the Sobral stay, stories, results,
and history see
\cite{rrfm1,rrfm2,crispino1,crispino2,crispino3,crispino5,joycemota}.

\begin{figure}[h]
{\includegraphics[scale=0.45]{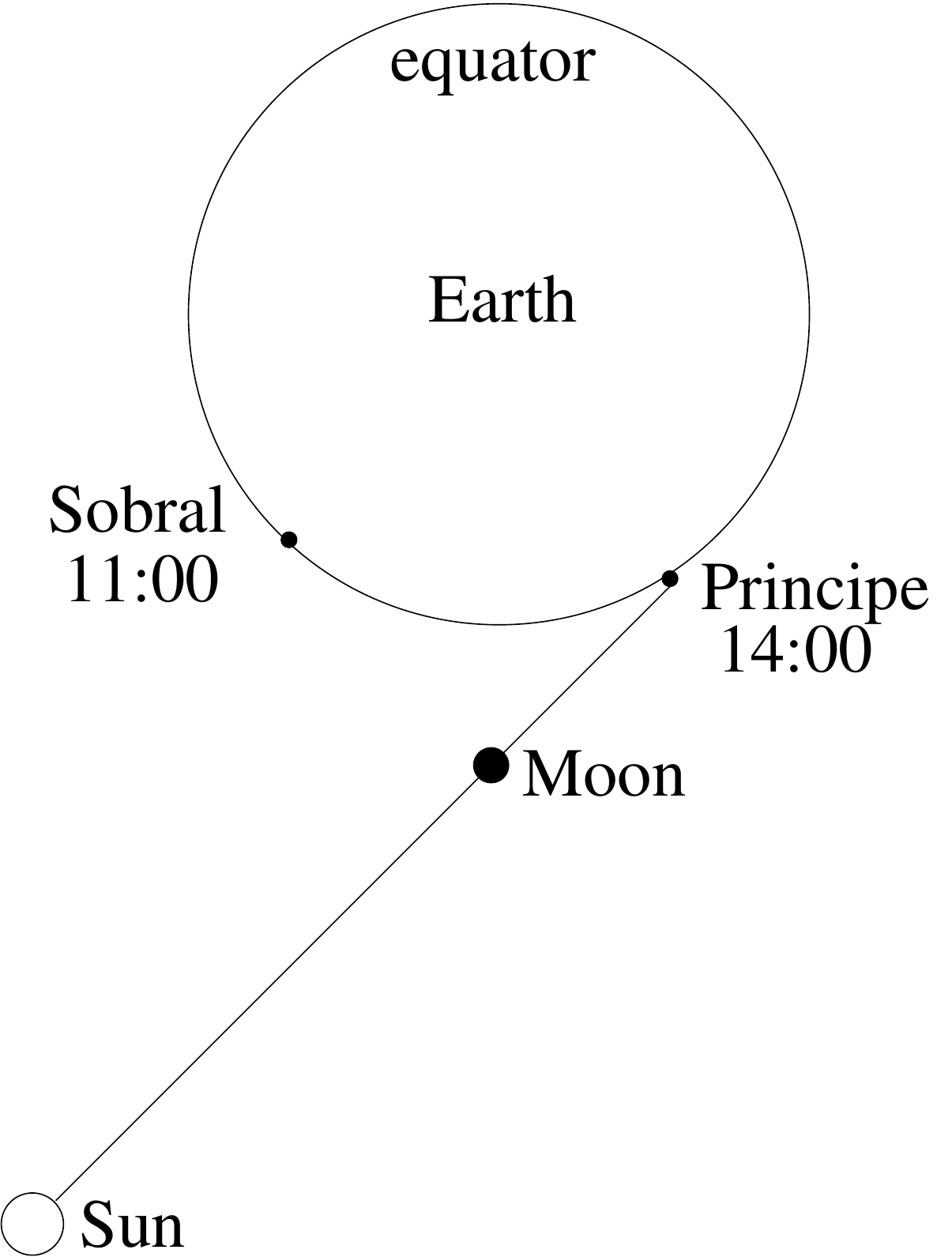}}
\caption{A view from the top of the Earth, Moon,
Sun system yielding an eclipse at 2:00pm local time
in Principe.}
\label{eclipseprincipe}
\end{figure}

In Principe, incredibly, there are no photos to tell the story in
pictures. Surely, Eddington and Cottingham forgot to take a camera
with them.  The eclipse in Principe was in the afternoon at 2:00pm,
see Fig.~\ref{eclipseprincipe} for a representation.  The hour
difference between Sobral and Principe is three hours, 
so taking that into
account, the Principe totality had a lag of about two hours in
relation to the totality in Sobral, as is clearly seen in the two
figures, Fig.~\ref{eclipsesobral} and Fig.~\ref{eclipseprincipe}.
These were old times, and certainly they would have liked to
communicate via teleconference or otherwise, but could not.

Although no photographs were taken in situ in Principe, 
the expedition is well documented by Eddington, 
both in the 1920 report \cite{report1920} and 
in his book Space, Time, and Gravitation, published 
also in 1920 \cite{eddstg1920}. In the book 
he states,
``On the day of the eclipse the weather was unfavourable.  When
totality began the dark disc of the Moon surrounded by the corona was
visible through cloud, much as the Moon often appears through cloud on
a night when no stars can be seen. There was nothing for it but to
carry out the arranged programme and hope for the best. One observer
was kept occupied changing the plates in rapid succession, whilst the
other gave the exposures of the required length with a screen held in
front of the object-glass to avoid shaking the telescope in any way.''
And immediately after he says in a poetic tone, ``We are conscious
only of the weird half-light of the landscape and the hush of nature,
broken by the calls of the observers, and beat of the metronome
ticking out the 302 seconds of totality''.
Yet, he writes \cite{eddstg1920}, 
``Sixteen photographs were obtained, with exposures ranging from 2 to
20 seconds.''
Eddington then after reducing 
one plate made the necessary measurements,
and not going mad after all, told 
Cottingham that he would not have to go alone
\cite{eddingtononcottigham}.
In Eddington's notebook it is written
\cite{allie}:
``June 3. We developed the photographs, 2 each night for 6 nights
after the eclipse, and I spent the whole day measuring. The
cloudy weather upset my plans, and I had 
to treat the measures in a different way from what I intended, 
consequently I have not been able 
to make any preliminary announcement of the result. 
But the one plate that I measure gave a result 
agreeing with Einstein.''
As Eddington referred to it, this was 
``the most exciting event I recall in my 
own connection with astronomy'' \cite{eddingtononcottigham}.
This echoes Einstein's words
about the discovery of the principle of equivalence  
that it ``was the happiest thought of my life'',
see e.g. \cite{pais1}.

Since the eclipse had been at 2:00pm in Principe,
in order that 
the Sun is rising at 6:00am
and the eclipse stars are in the same position
and can be seen before dawn,
one has to have a lag of 8 hours, i.e., 480 minutes, 
which is achieved in
(480 minutes)/(4 minutes/day)\,=\,120 days, i.e.
about 4 months.

So there was no point in staying that long, 
and after revealing some initial plates and reducing 
a few, they decided to
march back on June 12, transhipped at Lisbon, 
and arrived in Liverpool on July 14.
For details of the 
Principe stay, stories, results, and history 
see 
\cite{crawforscol,simoes1,simoes2,simoes3,crawfordgaz}.

\subsection{Data analysis}

As soon as the two expeditions arrived back in England the analysis of
the data started.  Dyson joined Crommelin and Davidson and they were
busy analyzing the Sobral plates at the Royal Greenwich Observatory.
Eddington in Cambridge was putting full effort in the analysis of the
Principe plates. The analyses by the two teams were independent.

\begin{figure}[h]
{\includegraphics[scale=0.30]{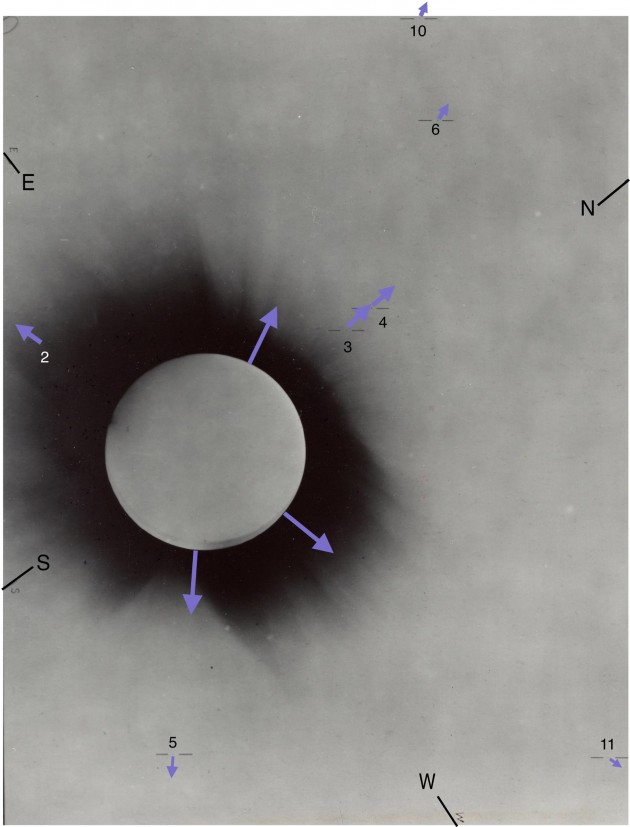}}
\caption{Photograph of the field of stars in the 
1919 eclipse by the 4 inch telescope in Sobral. Courtesy of 
The Royal Greenwich Observatory.}
\label{eclipse19194inchcortiesobral}
\end{figure}

In Sobral there were two telescopes.  There was the
small aperture, 4 inch telescope, handled by Crommelin, that was taken on
recommendation by Cortie as a back up, and that proved to be essential.  
There was
the main astrographic telescope maneuvered by Davidson, with
aperture of 13 inches that was reduced to 8 inches to get better
images.  The two telescopes can be seen clearly 
in Fig.~\ref{telescopesbritish}.
In
Sobral, with the 4 inch telescope 8 plates were taken with 7 stars
visible during the 5 minute and 13 seconds eclipse, see
Fig.~\ref{eclipse19194inchcortiesobral}.
With the astrographic telescope, 19 photographic plates were obtained
with 12 stars visible on them. Crommelin and Davidson 
then took photographs of the same field in the night sky 
seven weeks later and got the comparison night plates.

\begin{figure}[h]
{\includegraphics[scale=0.60]{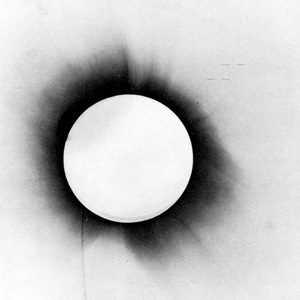}}
\caption{
Photograph of the field of stars in the 
1919 eclipse  in  Principe.  Some 
stars are marked.}
\label{eclipse1919astrographprincipe}
\end{figure}

In Principe, Eddington and Cottingham used only one telescope, an
astrographic telescope
similar to Sobral's with its aperture also reduced to 8 inches
coupled to a coelostat.  Here, 16 plates were obtained, of
which only 7 had stars, as the sky had a tenuous variable nebulosity
at the time of the eclipse, with 6 stars well visible, see 
Fig.~\ref{eclipse1919astrographprincipe} for a glimpse.
Eddington was able to develop immediately some photographic plates and
in reducing one managed to measure the full shift, confirming general
relativity. 
After returning to Cambridge he
analyzed the other plates and in the end only two plates were of worth
and 5 stars were used.  No comparison night plates were taken in
Principe and this obliged Eddington to solve the problem in a
different way.

Having the photographic plates with the stars in the environment of
the eclipsed Sun, one had now to quantify the shift in the stars'
positions due to the gravitational field of the Sun.  The procedure is
similar to the measurement of stellar parallaxes and star proper motions
in photographic plates and so was well-known.  The snag here is that
in an eclipse setting one is not at home, where all the equipment
works well and at prescribed times, and moreover, the eclipse runs for
a very short slot of time, of about five minutes.  Another difficulty
is that the maximum value for the displacement is for a star at the
Sun's rim. However, stars at the rim are rare and not bright enough to
be seen, the light from the corona shines them out.  Thus, the stars
observed in an eclipse are relatively far from the rim, say two times
or more further away, and the star's displacement has a lower value in
accordance to the hyperbolic $1/D$ law of Eq.~(\ref{hyperboliclaw}).
In order to see what is at stake, we note 
that the 1.75 arcseconds gravitational shift 
at the Sun's rim means, for the type of telescopes used,
about 0.03mm in the plate, and for a star at three solar radius
the gravitational shift is 0.58 arcseconds, about 
0.01mm in the plate.
Let us now look, in a nutshell, how both teams reduced 
the data to find the shift due to the gravitational field of the 
Sun. 

\begin{figure}[h]
{\includegraphics[scale=0.60]{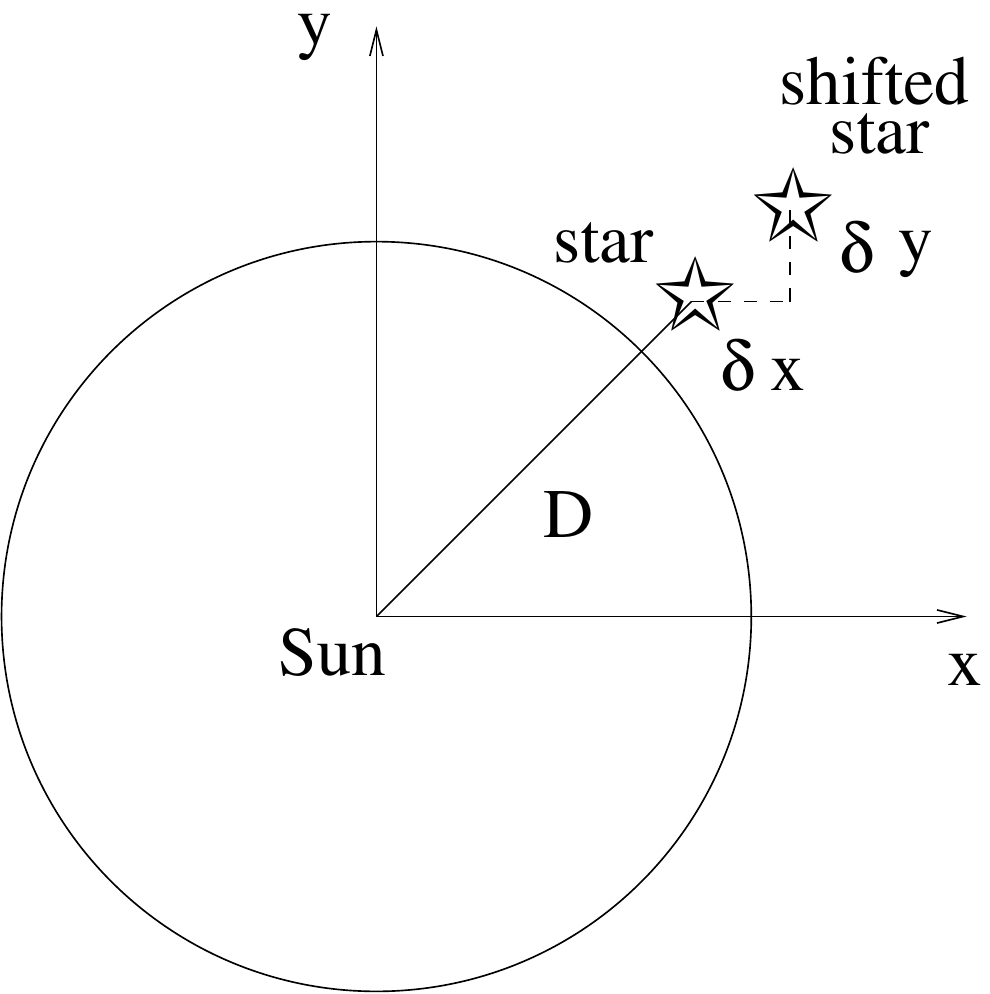}}
\caption{The position $(x,y)$ of the star 
and its measured shifted position 
$(x+\delta x,y+\delta y)$.}
\label{deltaxdeltayestrela}
\end{figure}

Suppose one has eclipse plates and comparison night plates.  
To find the shifts of the stars one has to put the two plates together.
In theory, one needs one star, one eclipse plate, and one
comparison night plate only. If all is perfect, one measures the
star's position in the comparison night plate, then measures the
star's position in the eclipse plate, subtracts, and gets immediately
the star's shift.
But, in practice, there are many more things involved.
There is mismanagement when putting the two plates together. 
The plates could be slightly translated and rotated in
relation to each other without notice. Also, more importantly there
could be a change of scale on one plate relatively to the other. This
change in scale comes from a change in the location of the focus on
the eclipse day and location of the focus on the comparison night day
that might appear due to variations of temperature or some other
factors.  One then makes a rectangular grid on the plates and assigns
an $x$ and $y$ position for each star in the comparison night plate.  
We follow closely, but not exactly,
the report of 1920 \cite{report1920}, see \cite{kluber}
and also \cite{moyer}.  
By comparing in both plates the positions of
the same star, the shifts $\delta x$ and $\delta y$ for that star can
be obtained, see Fig.~\ref{deltaxdeltayestrela}.

An equation for the displacement $\delta x$ and $\delta
y$ for each star can be written as,
\begin{equation}
\delta x=c+by+ax+\alpha\frac{x}{D^2}\,,
\label{shiftx}
\end{equation}
\begin{equation}
\delta y=f+dx+ey+\varepsilon\frac{y}{D^2}\,,
\label{shifty}
\end{equation}
where $c$ and $f$ are translation shifts that can come from some
anomaly in the clamping of the two plates, $b$ and $d$ are rotational
shifts that can also come from the clamping of the two plates, $a$ and
$e$ give the differences in scale value between the two plates, and
$\alpha$ and $\varepsilon$ yield the light deflection, with
$D=\sqrt{x^2 + y^2}$.  The form of the $\alpha$ and $\varepsilon$
terms comes from the $1/D$ dependence. Indeed, as the gravitational
shift is proportion to $1/D$, the respective shift in the $x$
direction is proportional to $\cos\theta/D=x/D^2$ and the respective
shift in the $y$ direction is proportional to $\sin\theta/D=y/D^2$,
where here $\theta$ is the angle between the $x$ direction and the
radial direction associated to the radius $D$.  One has four unknowns
for Eq.~(\ref{shiftx}) in $\delta x$, namely, $c$, $b$, $a$, 
$\alpha$, and
four unknowns for  Eq.~(\ref{shifty})
in $\delta y$, namely, $f$, $d$, $e$,
$\varepsilon$.  To solve the equations one needs a minimum of four
stars.  If one has $n$ stars with $n\geq4$ then the system is
overdetermined and one resorts to a least squares method.  Now, one
sees that $a$ and $\alpha$ scale both with $x$, and $e$ and
$\varepsilon$ scale both with $y$, which can bring difficulties in the
disentangling of $a$ and $\alpha$ and of $e$ and $\varepsilon$.
But, moreover, whereas the scaling increases  the shift 
relative to the center, the gravitational effect decreases the shift 
relative to the center, and at some distance $D$ they can be of
the same order, possibly increasing the difficulties.
On the other hand, since the 
gravitational effect obeys the $1/D$ law, 
stars far from the Sun, if there are any on the 
plates, have a negligible 
deflection and can in principle 
be used to set comparison points when 
the plates are matched. 
Further corrections that have to be taken into account
are
atmospheric refraction and aberration 
since the eclipse and comparison plates
are taken at different dates and times
of day.
Also, turbulence in Earth's atmosphere makes the stars 
scintillate and effectively produces random deflection 
of the light from the stars which can be nonneglible 
as compared to the gravitational effect. 
Nevertheless this effect is random and 
balances to zero for a sufficient number of stars.  
The analysis of the random 
and systematic errors in the measurements 
then follow standard procedures.

The Sobral team, now enlarged to be the Greenwich team with the
inclusion of Dyson to help in the data analysis, 
having eclipse plates,
comparison night plates, and an intermediary scale plate to insure
faithful comparison, did not bother with the scale problem.  They
simply solved Eqs.~(\ref{shiftx}) and~(\ref{shifty}) to get $c$, $f$,
$b$, $d$, $a$, $e$, and then finally the gravitational shift through
$\alpha$ and $\varepsilon$.

The Principe team, essentially Eddington, as Cottingham being an
instrument maker could not help in the analysis, did not get
comparison night plates, the eclipse had been at 2:00pm and the
eclipse star field would appear before dawn, on the night sky, four
months later only. Eddington managed to solve
the scale change problem because 
photographs of another star field were taken months earlier.
In fact, two different 
star fields in some January nights in
Oxford were photographed. The photographs 
were made  with the astrographic telescope that went to the eclipse, which
belonged to Oxford.  One star field that was photographed in Oxford was
the star field that would pop in the eclipse day in Principe, 
called the
comparison night field.  The other star field that was photographed in
Oxford was a given chosen star field, the check field.  This check
field was then also photographed in Principe nights in May.  
All this trouble  
was to take precautions against 
any change on the lens on the trip
and to take care of systematic errors 
that could derive from 
the distinct conditions 
at Oxford and Principe.
In the end the check 
field plates were essential in the analysis.
This was because there were no sufficient
stars to provide the data 
necessary to 
find the constants 
in Eqs.~(\ref{shiftx}) and~(\ref{shifty}) and 
pick up in the end
the pursued $\alpha$ and $\varepsilon$.
The check plates would thus function as 
appropriate for 
systematic error determinations 
as well as to settle 
the constants 
in Eqs.~(\ref{shiftx}) and~(\ref{shifty}).
Indeed, by
applying Eqs.~(\ref{shiftx}) and~(\ref{shifty}) to the two check field
plates, without the $\alpha$ and $\varepsilon$ terms as there was no
gravitational deflection for the check field, Eddington managed to get $a$
and $e$ for the scale change in an independent manner.  He then
compared the eclipse field in Principe with the comparison night field
of Oxford, knowing beforehand the scale change factors $a$ and $e$.
Going then back to  Eqs.~(\ref{shiftx}) and~(\ref{shifty})
with $a$ and $e$ known,
and correcting for refraction and aberration, he could
determine $\alpha$ and $\varepsilon$
in an ingenious manner.  
This method, 
using several complementary layers 
to obtain the final result, was 
invented by
Eddington by necessity 
and proved to be useful and used in other eclipses.

\subsection{Results}

After the return to England of the British teams, people interested
in general relativity and positional astronomy started to get
impatient.  Certainly, the most anxious of all was Einstein.  In
September he inquired Dutch colleagues if there were 
developments from the
eclipse expeditions.  
\begin{figure}[h]
{\includegraphics[scale=0.40]{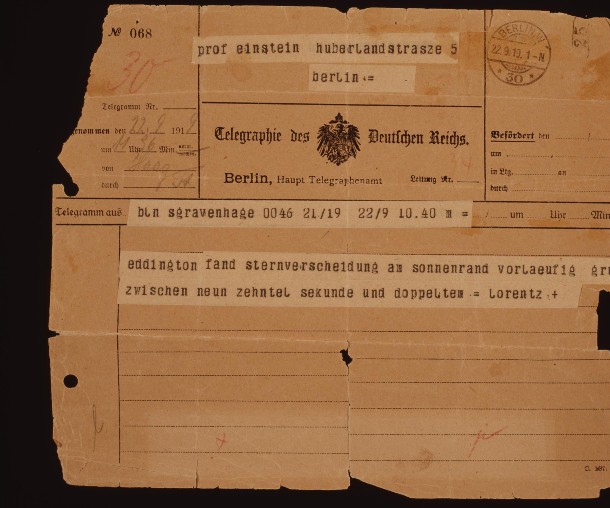}}
\caption{Lorentz telegram to Einstein in September 22, 1919, stating
``Eddington found stellar shift at solar limb, tentative value between
nine-tenths of a second and twice that.''}
\label{telegram}
\end{figure}
Then, just after, Lorentz
hearing that Eddington had claimed in a meeting that the results
indicated some amount of gravitational light deflection but could not
tell yet the precise values, immediately sent a telegram to Einstein
with the good news, see Fig.~\ref{telegram}.  Einstein jubilated. 
In that same day he wrote to his mother: ``Good news
today. H. Lorentz sent me a telegram saying that the British
expeditions confirmed definitely the light deflection by the Sun.''
Sometime later he then states his famous phrase when the
student of philosophy Ilse Schneider asks what he would say
if the eclipse 
results were otherwise, not confirming his predictions: ``I
would have to be sorry for the dear Lord. The theory is correct'', 
see e.g. \cite{clarkebio}.

The analysis at 
the
Royal Greenwich Observatory and Cambridge, made along the lines already
described, was performed from August to October.  Eventually, on
Thursday, 
November 6, 1919, in a joint meeting of the Royal Society and the
Royal Astronomical Society, at the Royal Society with a room
completely crowded, the results of the two expeditions were announced.
J.~J.~Thomson, the man of the electron and the
President of the Royal Society, opened the
session.

The word was then passed to Dyson and Crommelin as representants
of the Sobral expedition and of the Royal Greenwich 
Observatory analysis.
They stated that the 4 inch telescope had 7 very good stars and the
result for the angle of deflection  $\delta$ by the gravitational
field of the
Sun at its rim was 
\begin{equation} 
\delta=1.98\pm 0.12\,{\rm arcseconds}.
\end{equation} 
Dyson then declared about the plates and the reduction
process \cite{jjthomson}:
``I am prepared to say that 
there can be no doubt that they
confirm Einstein's prediction. A very definite result has been
obtained that light is deflected in accordance with Einstein's law of
gravitation.''
Dyson and Crommelin
also said that the 13 inch astrographic telescope system
did not work
properly, the stars' images were diffused and to explain this
there were two possibilities,
either the telescope was out of focus or it simply blurred
the images.
This had happened 
probably because the coelostat heated and
in consequence 
mal-functioned.
If it was assumed that there was a change of
focus due to the mal-function
and the eclipse plates were used
to determine the scale,
in the way mentioned above, then the value 
for the deflection was 0.98 arcseconds. 
If it was assumed that the stars were simply
blurred with no change of scale and 
no mal-function of the
telescope system then the value was 1.40 arcseconds.
Thus, being hard to trust
the astrographic telescope, 
the Royal Greenwich Observatory team decided to 
discard its results.

Then Eddington spoke on the
Principe outcome \cite{jjthomson}. 
He said that the 
Principe 13 inch astrographic telescope
and coelostat functioned 
well, admittedly because the temperature 
was not too high due to the cloudy veil.
He stated that there were five good stars 
and the
result for the angle of deflection $\delta$ 
by the gravitational field of the
Sun at its rim was 
\begin{equation}
\delta=1.61\pm0.30\,{\rm arcseconds}\,.
\end{equation}
He then interpreted the results.
For the
half effect Newtonian gravitation
would be the correct theory.
The full effect
had been obtained and so
gravitation obeys the law
proposed by Einstein. 
He mentioned that
the new law
had already incorporated
Mercury's perihelion precession 
and spoke of the possibility
of detecting in the future the gravitational
redshift in the surface of the Sun predicted by the
law. 

Then J. J. Thomson
raised to the occasion and said \cite{jjthomson}:, ``This is the most 
important
result obtained in connection with the theory of gravitation since
Newton's day, and it is fitting that it should be announced at a
meeting of the Society so closely connected with him.''  The moment
was echoed by the philosopher Whitehead some years later musing about
the meeting. He wrote \cite{whitehead}, ``in the background the
picture of Newton to remind us that the greatest of scientific
generalisations was now, after more than two centuries, to receive its
first modification.'' And then, ``a great adventure in thought had at
length come safe to shore.''

All the details of the data analysis and results
of the two expeditions
were then written down in the report of January 1, 1920
\cite{report1920}
with Dyson, Eddington, and Davidson as authors. 
Why Crommelin was not an author in the report 
has not been
explained. He led the Sobral
expedition, he was in charge on the day of the 4 inch telescope
that yielded the good data, he
was in the data reduction process, he talked
in the historical joint meeting of November 6, he published papers
on the eclipse before and after the eclipse,
but in the most important
paper, the 1920 report, he is not an author.
In concluding, 
the report states in the end that 
\cite{report1920}
``In summarising the results
of the two expeditions, the greatest weight must be attached to those
obtained with the 4-inch lens at Sobral''.
This judgement is certainly because the error of the results 
in Sobral were $\pm0.12\,$ arcseconds 
whereas in Principe
they were $\pm0.30$, two and a half times larger, and so
admittedly less reliable.

It has often been said that the conclusion,
that Einstein's theory has been proved on the 
1919 eclipse observations 
by the two expeditions, 
has been taken on shaky data 
and shaky grounds, see, e.g., \cite{earman}. 
The last figure of the 1920 report \cite{report1920} plots
gravitational shift due to light 
deflection versus the inverse distance 
from the center of the Sun $1/D$, with $0<1/D<1/R$,
where $R$ is the Sun's radius. Three straight lines 
popping from the origin ($1/D=0$, i.e., $D=\infty$)
are given
in that figure. The lower straight line, the less 
inclined dotted line, is for the 
Newtonian prediction, 
the middle heavy line is for 
the Einstein prediction, 
and the upper straight line is for 
the observations carried on the two sites. 
This upper straight line, does not match
the Einstein prediction,  yields in 
fact values for the deflection 
along the $1/D$ abscissa that
are slightly higher than the Einstein prediction.
Thus, the maximum that one could say was 
that (i) no deflection is ruled out, 
(ii) Newtonian gravitation with light 
coupling to it is ruled out, 
and (iii) general relativity might be correct, 
which
could only be tested by future observations.
The authors nevertheless claim \cite{report1920},
``Thus the results of the expeditions to
Sobral and Principe can leave little doubt that a deflection of light
takes place in the neighbourhood of the sun and that it is of the
amount demanded by Einstein's generalised theory of relativity, as
attributable to the sun's gravitational field.''

So, on what evidence and with what arguments could one say that
Einstein's theory had been proved in 1919? There are several. First,
Newtonian gravitation could no more be upheld on fundamental
principles. By then it was known to some that the world was
relativistic and the velocity of propagation of any signal was finite
contradicting on its face Newton's law of gravitation where the
propagation of the gravitational field is instantaneous, and thus
its speed of propagation is
infinite.  
Moreover the concept of mass was no more
invariant, mass of a particle would 
change depending on the observer bringing
further complications to Newton's laws.
Second, Einstein had mounted a theory based on solid
principles, namely, special relativity principles, the equivalence
principle, the covariance principle, 
and the metric principle which states that the world is
described by a spacetime metric, as Minkowski prescribed.  Third, as
soon as the theory was about to be ready, Einstein calculated
Mercury's perihelion precession and found 43 arcseconds per century,
the amount that was lacking in Newtonian gravitation, a problem
without solution for decades.  This is astonishing, as the
conceptually complex theory, as general relativity is, when applied to
the concrete problem of the trajectory of planets, has in its kernel
and essence, the precise amount required by the data. However,
wonderful as it was, this precession was an a posteriori confirmation,
not an apriori prediction, and thus confirmation of new 
general relativistic predictions
were necessary.  Fourth, and finally, there was indeed the general
relativistic definite prediction of 1.75 arcseconds for the light
deflection by the gravitational field of the Sun.

Given that the eclipse data was surely on the full deflection side, 
it was clear
that general relativity had been favored by a great 
margin, 
and that the other two possibilities, no deflection
for the case light does not interact
with gravitation, or 
half deflection for the case Newtonian gravitation was the
correct theory and light coupled to it, had been ruled out,
see also \cite{kennefick}.
Moreover, given the other three points 
raised above, it was obvious, especially for a theorist 
that Eddington was, that 
the measurements had confirmed general relativity. 
Or in other words, it was highly probable 
given the four points above that general relativity
was correct. 
One could think on extensions of general relativity
which could include electromagnetism 
on its fundamental level, as Eddington was already 
thinking at this time, drawing on  work
by the German physicist and mathematician 
Hermann Weyl, but these extensions in principle
would not affect the 
general relativistic light deflection
result at least at some
zeroth order level.
Therefore Dyson and Eddington 
were right in betting without further ado 
on general relativity. 
With humor it is said that
the 1919
data did not prove that Einstein was right,
Dyson and Eddington did.
It is also frequently quoted that 
Eddington  said that he did not need 
to go to Principe to prove general relativity
since he 
was fully convinced of the truth of the theory
\cite{chandrasekhar}.

Interestingly, the analysis of the plates and data from Sobral was
repeated with all the new modern available techniques by Harvey from
the 
Royal Greenwich Observatory in 1979 \cite{harvey}, which 
having passed unnoticed to Hawking
\cite{hawking}, elicited a comment by Wayman and Murray
\cite{waymanmurray}.  The reanalysis revealed for the 4 inch object
glass the value $1.90\pm0.11\,$arcseconds, and for the 13 inch object
glass the value $1.55\pm0.34\,$arcseconds.  The 4 inch telescope
yielded trivially about the same results with less error, the the 13
inch telescope had, after all, utilizable data that gave results in
harmony with the other telescopes. All is well that ends well.
Unfortunately, most of Principe plates and Eddington's reductions
inexplicably disappeared from the Observatories in Cambridge, probably
into the litter, after Eddington's death.

\subsection{History}

The war had finished a year before, and there were still bitter
feelings among the belligerent countries. This confirmation of the
light deflection had international relevance.  German science was
being shun by the British and this event gave a hope to stop the ban
on the Germans as was the desire of Eddington, a pacifist in
character.  The scientific cooperation was an example for all the
world to see, that British science represented by the two eclipse
expedition teams had confirmed a completely new and profound theory of
gravitation proposed by a scientist on the other side. On the top
of the scientific achievement many appreciated this peace effort
accomplishment.

History had been done.  The newspapers knew that and announced with
jubilation the achievement of the observations which, in turn,
spurred a huge enthusiasm from the public at large.

The major journals of England immediately reverberated 
the accomplishment
of its scientists. General relativity was
held as the new theory of gravitation and of the 
Universe. The Times of London, 
revealed the results in November 7, one day 
after the joint meeting of the two royal societies, writing
shortly afterwards an article with the title 
``Revolution in science: 
The ideals of Aristotle, Euclides, and Newton, that are
the basis of our conceptions, do not correspond 
to what can be observed in the structure of the universe''.
In the
USA, the New York Times announced the results in November 10, the O
Jornal from Rio de Janeiro, in November 12, and the O Jornal O Século,
from Lisbon, gave the information in November 15. The Illustrirte
Zeitung from Berlin wrote in December 14, ``A new celebrity in the
world history: Albert Einstein, whose investigations lead to a
complete revision of our concepts about Nature and are at the same
level of those of Copernicus, Kepler, and Newton''.  The German
journalist forgot to include Galileu but that is another matter.

Einstein was acclaimed instantaneously.
Up to then he was known as an outstanding 
theoretician
within a circle of physicists working in 
areas related to his works. The announcement of 
the 1919 eclipse results catapulted him to
celebrity.
Indeed, Einstein's biographies \cite{clarkebio,pais1,pais2}
highlight the results of
the Sobral and Principe expeditions.
For instance the biography by Clarke
\cite{clarkebio} opens a chapter of the biography stating ``In the
morning of November 7, 1919, Einstein woke up in Berlin as a famous
man.''  
From that day onwards he became a world
figure and started to be invited
to all places around the planet.

As an example of his voyages, in March 11, 1925, on the way to Brazil,
Argentina, and Uruguay, the ship he was travelling in stopped in
Lisbon for two days.  
He visited the Castle of São Jorge, a fascinating site whose
initial fortifications date back to the first century, and the
Monastery of Jerónimos, a wonderful late Gothic Manueline style
construction of 1500, but he was really impressed with the varinas,
fisher women in downtown 
Lisbon,
annotating in his onboard diary, ``A fisher woman selling fish,
photographed with a fish basket, proud gesture, naughty''. 
In Lisbon, nobody noticed his passage, in spite of
being very famous by now; on the top of the 1919 eclipse results hype,
he had already received the 1921 Nobel Prize in Physics
\cite{fitas}.  Perhaps this
can be justified by the political turmoil the country was in.  Mira
Fernandes, a leading Portuguese mathematical physicist, working in
Lisbon at Instituto Superior Técnico, was at this time starting to get
interested in general relativity and in the ideas of unification of
gravitation and electromagnetism \cite{jpsl1,jpsl2,jpsl3}. 
He was most probably
not aware of Einstein's stay in the capital, otherwise he would have 
invited Einstein to give a talk or
at least to meet him.  He compensated this failure somehow
by, under his suggestion,
having Einstein together with Levi-Civita, elected as foreigner
correspondents of the Lisbon Academy of Sciences in 1932, a year
before Einstein  left Berlin to Princeton. 
Interestingly, on the day following this election, March 18, 1932,
the newspaper Pittsburgh Press announced
this ceremony.
Also, Santos Lucas, a
professor of mathematics in the Faculty of Sciences of Lisbon, would
have liked to meet Einstein in his passage in 1925. He had
dedicated a whole semester to
a course in general relativity, where the light deflection
phenomenon is treated with rigor, surely one of the first courses in
general relativity in the world \cite{santoslucas},
see also \cite{jpsl3}.  Much
later, in 1946, António Gião, a physicist based in Lisbon, with works
in general relativistic cosmological solutions and unification
schemes, corresponded with Einstein, a fact he was very proud of
\cite{jpsl3}.
After leaving Lisbon the ship where Einstein was
headed for Rio de Janeiro 
where he arrived in March 21, 1925,
for a three day visit. He then
proceeded to Buenos Aires to give a series of lectures
in the university, continued to Montevideo, 
and passed again in Rio, where he now delivered
lectures in several places, visited Henrique Morize
in Observatorio Nacional in May 9,  
and socialized 
with high authorities.
Finally, he returned to Hamburg on May 12. Einstein's visit in Rio
is well documented, see e.g.,
\cite{ildeu,tomasquim}.

As for Eddington he also became a figure not only in scientific
circles, but in larger public circles, from 1919 onward.
The 1919 events, of which he was the
exponent, transformed the scene, signaling to the world the
importance of general relativity as the correct theory of
gravitation. How important was luck in all of this?
As usual luck plays a part. Of course, sooner 
or later, general relativity would
be vindicated, some eclipse would show 
the correctness of general relativity, 
but this one was spectacular and surely Eddington played the major
role in its magnificentness. His influence was vast and profound.
For instance, Dirac was so impressed by the inspiring
1919 eclipse test that he moved 
from Bristol to Cambridge to study relativity
with Cunningham but ended up being a student
of Fowler to work on quantum mechanics
\cite{moyer2}, see also \cite{moyer}.

\begin{figure}[h]
{\includegraphics[scale=1.6]{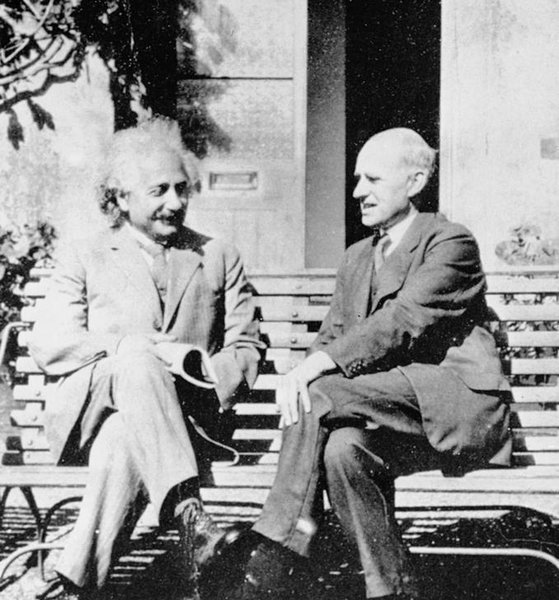}}
\caption{Einstein and Eddington in front
of the main building of the Observatories, in
Madingley Road, Cambridge, in June 1930.}
\label{eddingtoneinstein}
\end{figure}

In 1930 Einstein finally went to Cambridge.  He and Eddington
corresponded along the years after the eclipse confirmation and
Eddington invited Einstein to visit Cambridge.  Fortunately a
photograph of both was taken, see Fig.~\ref{eddingtoneinstein}.
It says all, two happy men posing for posterity.

In the 1920s through to the 1930s general relativity was taken by
mathematical works.  Perhaps, the great exception is the work of
Oppenheimer and Snyder of 1939 that discovered black holes by proving
mathematically their existence
\cite{os}.  Oppenheimer was aware of his discovery
and that black holes were real \cite{thornebhs}, but couldn't care
less, he understood it as a minor contribution \cite{dyson}, an
amazing fact on several grounds. This year
we are thus also celebrating eighty years of black holes. 
The name black hole itself appeared
only later through the hand of Wheeler
\cite{herdeirolemos}. Then after the second world war, in the 1950s,
a renaissance of general relativity back into physics took place
\cite{eisenstaedt}, with
Wheeler and his group in Princeton where at its beginning Einstein was
an inspirational figure still around, with Hoyle,
Sciama, Penrose, Hawking,
and Carter in Cambridge, surely in Eddington's tradition, and with
Zel'dovich and his group in Moscow \cite{israel}.
General relativity is now taught in 
most, if not all, undergraduate courses in physics.
Books in general relativity abound 
and the student can learn
relatively quickly this discipline. 
To cite some, there is the book of
Eddington \cite{eddmathematicalrelativity}, a masterpiece,
there is the book by Adler, Bazin, and Schiffer \cite{adler}, 
written in 1965 and still 
a marvelous book in its modernity, 
there is the book by Weinberg \cite{weinberg},
of 1972, with
the calculations done in a beautiful way and with
an emphasis on cosmology, 
carefully treating all its mathematical and physical 
aspects,
there is the book by Misner, Thorne, and Wheeler,
Gravitation \cite{mtw}, written in 1973,
an extraordinary treatise up-to-date in its approach
and encompassing subjects, 
there is the book by Wald 
with advanced topics \cite{wald},
there is the book by D'Inverno \cite{dinverno}, 
that made general relativity a very
easy subject to teach and learn, 
and there are others excellent 
books, e.g., 
\cite{krasinski,ryder}.

Nowadays, general relativity is basic to understand the whole
cosmos and its functioning,
from cosmology to black holes, to gravitational
waves, to gravitational lensing, to other 
disciplines, all are products of general relativity.

\section{Post eclipse and the future}

There is a long list of expeditions to eclipses to test the light
deflection prediction by general relativity following the 1919
eclipse.  We name a few.  Campbell, after two failed attempts was
desperate to get results. So, he took from the Lick Observatory all
the necessary paraphernalia with 35 tons of equipment, to land in
Wallal beach, in west Australia.  It was September 21, 1922, and the
weather was fine when the Moon darkened the Sun. Later he would report
that the measurements gave $1.72\pm0.15\,$arcseconds 
for the deflection angle,
confirming general relativity's prediction 
right on target. It has been 
considered the best determination of the light deflection in
an eclipse of the Sun.  Other expeditions went to observe this
eclipse, some were rain washed others got good results
\cite{kluber}. 
Freundlich, who together with Einstein had started all
this business, had had permanent bad luck. In 1914 he was made
prisoner in Crimea, in 1922 he chose Christmas Island, the wrong
site, as it rained on the day, in 1923 he got rain in Mexico, in 1926
in Sumatra once again it rained, and finally in 1929 also in Sumatra
the weather was fine, the measurement gave a deflection considerably
greater then general relativity's prediction, 2.2 arcseconds 
\cite{kluber},
and nobody else
believed in it.  A team from Austin, Texas, that included the
relativists Bryce DeWitt, Cecile DeWitt, and Richard Matzner, organized
an expedition to the Mauritania desert, for the eclipse of June 30,
1973. They witnessed a dust storm just before the eclipse, but even so
they managed to find from the emulsion plates a light deflection of
$1.66\pm0.19\,$arcseconds \cite{mauritaniatexas},
see also \cite{littmann}.  
In present times, amateur astronomers with high expertise
on telescopes and optics, in possession of CCD cameras, can test the
light deflection from the Sun, as was done by Bruns in the August 21,
2017, eclipse in Wyoming, USA, obtaining 1.7512 arcseconds for the
deflection \cite{bruns}.

The light deflection tests have improved in many directions.
One is in gravitational lensing.
Since a gravitational 
field deflects light it can 
act as a gravitational lens, in which case
the light rays from behind
the gravitational field source
converge at the observational point.
\begin{figure}[h]
{\includegraphics[scale=0.35]{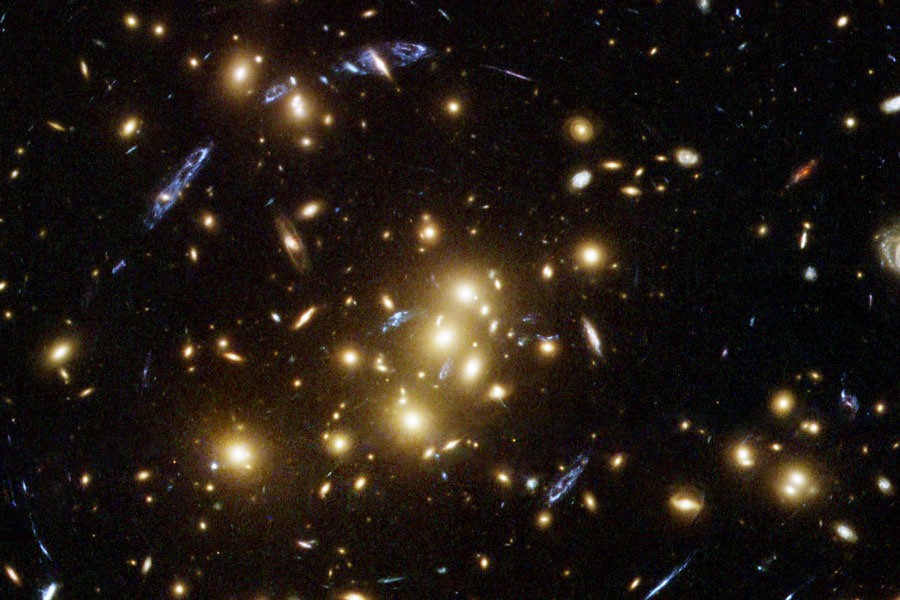}}
\caption{Wonderful light deflection.
A blue galaxy appearing several times along
a ring is being lensed by a cluster of galaxies. 
Courtesy of the 
Hubble Space Telescope - NASA.}
\label{lentegravitational}
\end{figure}
Light deflection from stars behind the Sun
does not produce a lens at Earth, the light rays are not 
enough deflected to form a lens. Lensing
is thus a special, though abundant,
case of light deflection.
Einstein in 1913 thought
that stars themselves 
might be gravitational lenses of other stars, 
but found the effect negligible and opted 
for not publishing it until much later
when in 1936 he was pressed to disclose these calculations.
Just after,
Zwicky, a Swiss astronomer working in Caltech
that had several other influential insights, 
showed that the lensing effect was non
negligible for galaxies and clusters of galaxies
acting as gravitational deflectors 
of light coming behind them from, e.g., another galaxy
or another cluster of galaxies.  
The effect was confirmed in 1979 when two 
quasars with the same properties  were identified
as one single quasar for which its light was 
suffering lensing from an intervening galaxy.
Today, these lensing effects are detected all over, see
Fig.~\ref{lentegravitational},
where an example of an 
Einstein ring is displayed in full glory.
Gravitational lensing serves to measure 
the mass of the galaxy responsible for the effect
as well as giving a measure for
the dark energy that is responsible for
the acceleration of the Universe
\cite{schneider}.
Gravitational lensing in cosmology
will be done by the ESA mission
Euclid and the NASA mission WFIRST, both of them designed to set
constraints on the physics of the dark energy \cite{tereno}.
Currently, black holes and accretion disks provide 
strong field light deflection and lensing, 
called black hole shadows, as
the Event Horizon Telescope has showed, see e.g.
\cite{jpslemosherdeirocardosogazeta}.

Another direction in which gravitational
light deflection is proving important is
optical astrometry. The ESA
satellites, Hipparcus
and Gaia, measured 
star positions up to milliarcsecond
and microarcsecond precision, respectively.
Gaia, still in operation, is 
measuring the positions of billions of stars.
These positions have to 
to be corrected
for the light deflection
from the Sun, Jupiter, and other planets,
as the measurements
are so precise that without the corrections
the star sky mapping would be incorrect,
see, e.g., \cite{perryman,perrymanpaper}.

Yet another direction
for light deflection 
is
with the use of radio telescopes which can measure positions with
great precision.  Shapiro, an American astrophysicist that proposed
the gravitational time delay, an effect that bears some relation to
gravitational light deflection, also had the idea of testing
gravitational deflection, but using radio waves and radio telescopes
instead of light, by bouncing those waves back off an inner planet of
the solar system when the Sun and the planet are in a correct
alignment with the Earth \cite{shapiro}.  Radio wave interferometry to
test light deflection proved to be much more effective by the use of
quasars, objects at cosmological distances, many emit strongly
in the radio. The quasars are fixed in the celestial sphere and some
of them are in the plane of the ecliptic. When, as seen from the
Earth, the Sun passes in front of such a quasar its angular separation
from another quasar in a nearby place in the ecliptic changes due to
the gravitational light deflection effect. A value that has been
obtained is $1.75$ arcseconds 
with a very small residual error \cite{shapiro2,
fomalont}, again in perfect accord with general relativity.
The precision on the position of radio quasars
is so great that
it is now possible to measure the deflection due to the Sun
of the radio waves emitted from quasars
that are reasonably far from the Sun in the celestial sphere
and even to measure the deflection due to the
stars in our Galaxy
of radio waves emitted from those quasars
\cite{sazhin}.

The initial criticisms against general relativity
soon died away,
and instead a few physicists started to consistently build, after
1919, new theories of gravitation
grounded in great part on the ideas of general relativity. Now, there is
a plethora of such theories, and due to all the high precision in the
optical and the radio telescopes, as well as due to gravitational wave
detection, one can now test alternative theories to general
relativity.  All these alternative or modified theories of gravitation
can be parameterized through the parameterized post-Newtonian, or PPN,
formalism \cite{will1}. And so slowly and surely we will know
at which level general relativity will go astray.

We end by coming back to the start. Mercury's precession,
gravitational redshift, gravitational light deflection, and the
gravitational time delay, were tests possible to do within the solar
system and all have all confirmed general relativity.  The gravitational
redshift has found a direct technological application in the GPS, as
to work properly with all its clocks synchronized the effect has to be
taken into account.  Gravitational light deflection has as a
particular case gravitational lensing which serves to measure masses
of the intervening objects and as a tool for cosmology.  Cosmology,
the study of the Universe as a whole, has showed that the Universe is
in accelerated expansion.  Fundamental theories, that started with
tentative unifications of gravitation and electromagnetism, by Weyl,
Eddington, and Einstein, continued to be pursued, now they are called
theories of everything and try to unify the four fundamental fields in
a unique quantum scheme.  Black holes, the geometrical
object par excellence of general relativity, can
now be seen through spectacular light deflection, called black hole
shadows, and also provide a tool to probe quantum gravity through the
Hawking radiation effect.  Gravitational waves, or spacetime waves,
predicted by Einstein in 1916, have been detected directly in 2015 by
LIGO and were generated by the collision of two black holes.  In
brief, general relativity has so far provided a great adventure that
has been thoroughly enjoyable.  Is the end in sight? Not yet.

\section{One hundred years after: The 2019 celebrations}

The year 2019 marks the one hundred years of the light deflection
observations and consequent confirmation of general relativity.
Given the historical character of this date several
celebrations have been organized.

In Sobral there was a scientific conference and a
public event from May 26 to May 31, 2019,
that certainly lived up to the importance
of the discovery, see \cite{ildeugazeta}.

In Principe there was a scientific conference 
``From Einstein and Eddington to LIGO: 100 years of 
gravitational light deflection'' 
with the main aim of
celebrating such an important date with worldwide experts to reflect
on the legacy left by Einstein and Eddington and to discuss the
subsequent startling developments in the fields of astrophysics and
gravitation, namely, black holes, gravitational waves, gravitational
lensing, and cosmology.
The meeting was held from May 26 to May 30, 2019,
the webpage of it is 
https://science.esundy.tecnico.ulisboa.pt/en/, 
and the organizers were 
Vitor Cardoso, Carlos Herdeiro, and the author
of this article. 
There was a public event at Roça Sundy on May 30, exactly
one hundred years after the eclipse, see 
http://esundy.org/index.php/en/homepage/ and 
\cite{latas}.
Sobral and Principe got together on May 29 on a
videoconference where scientific and political
personalities congratulated each other for this
special moment. A special number of the 
Portuguese 
journal Gazeta de Física 
has been issued to celebrate the events in Sobral
and Principe
\cite{jpslfitascrawford}.

Finally, in London there was a public evening event
on November 6, 2019, organized by the Royal Astronomical
Society to celebrate the one hundred years of the
announcement of the results of the light deflection by Dyson,
Crommelin, Davidson, and Eddington that confirmed Einstein's theory of
gravitation.

\begin{acknowledgments}

I thank Nelson Studart for having invited me to write this article to
the special number commemorating the 100 years of the eclipse in the
prestigious journal Revista Brasileira do Ensino de Física of which he
is editor.

I thank Júlio Fabris from Vitória-ES, Nelson Pinto Neto and Santiago
Perez Bergliaffa from Rio de Janeiro, and Vilson Zanchin from São
Paulo, for, on the understanding of the significant importance of the
event, having organized on short notice a great scientific conference
on the eclipse and general relativity in Sobral, which was the
symmetric part of the scientific conference organized by us in
Principe, and for very many conversations on general relativity and
cosmology.  I thank Ildeu Castro Moreira, President of the Sociedade
Brasileira para o Progresso da Ciência SBPC and organizer of the main
public event in Sobral, for all the help in the setting of the major
videoconference across the Atlantic between Sobral and Principe, on
May 29, 2019, eclipse times, celebrating the 100 years on the spot,
that involved scientists and politicians from both sites, and also for
conversations on the eclipse.
I thank Emerson Almeida, of University of Vale do Acaraú in Sobral,
for, in February 2019, showing me all the important places where the
expedition teams were located and made the observations, and for
inviting me to give a colloquium on the eclipse in the university with
a participation of more than 200 students.  I thank Luis Carlos
Crispino of Belém do Pará for many conversations on the Sobral
eclipse, in particular for pressing me, and for that matter all the
world, that Sobral of Crommelin, Davidson, and Dyson, has to have the
same importance in the news
as Principe of Eddington. I thank Claudio Bastos of
Observatorio Nacional of Rio de Janeiro for many conversations on
astronomy and astrophysics and on the eclipse.

I thank all my group CENTRA at IST Lisbon, for the support in the
organization of the Principe scientific conference ``From Einstein and
Eddington to LIGO: 100 years of gravitational light deflection'',
realized in May 26-30, 2019, Sérgio Almeida of CENTRA for the help in
the administration of the conference's webpage, and Dulce Conceição of
CENTRA for handling all the administrative processes for the
conference. I thank our colleague Luis Viseu Melo,
in charge of the IST finances, for simplifying the necessary 
internal procedures.
I thank the conference coorganizers
Vitor Cardoso and Carlos Herdeiro
for all the good atmosphere created in the preparation of this event.
I thank Phillipe Moreau and Beatriz Geraldes of HBD for
assisting with great sympathy in the logistics of the conference in
the resort Roça Bom Bom, and the manager Nuno Santos for all the help
at the resort.  I thank Joana Latas of NUCLIO for the help she gave in the
arrangements for a smooth interface between the scientific activities in
Roça Bom Bom and the general activities in Roça Sundy and in providing
infrastructure for the videoconference between Sobral and Principe
on May 29, 2019, eclipse times.

I thank André Ferreira Freitas from São 
Tomé for inviting
me to give a seminar by teleconference on May 29, 2018, 
for the 99 years of the eclipse, for the students of the Portuguese 
School of São 
Tomé e Principe.
I thank Paulo Crawford and Ana Simões  of
Faculdade de Ciências, Lisbon,
for many conversations on the history of the eclipse 
and, together with Augusto Fitas,
for the teamwork in the preparation of the
special number of Gazeta de Física 
dedicated to Einstein,
Eddington, and the eclipse.
I thank Ismael Tereno of Faculdade de Ciências, Lisbon, for many
conversations on the physics and astronomy of the 1919 eclipse and on
gravitational lensing.  I thank Ana Mourão and Ilidio Lopes of CENTRA
and IST, Lisbon, for allowing me to speak on the 1919 eclipse and
general relativistic light deflection in the stimulating XXIX
Astronomy and Astrophysics National Meeting in Lisbon in September
2019, and Amaro Rica da Silva of CENTRA and IST, Lisbon, for the many
conversations on eclipses and light deflection, in particular in
helping in the light deflection drawing of Fig.~\ref{lightdeflection}.
I thank David Hilditch of CENTRA and IST for 
conversations on the eclipse. I thank Diogo Bragança, MSc student 
at CENTRA and IST three years ago, now 
in Stanford finishing his PhD, for the careful reading of the
manuscript.

I thank Donald Lynden-Bell, my PhD supervisor in the Institute of
Astronomy, Cambridge, for the many enlightening conversations on
physics and astrophysics, and on Eddington. He was a greater admirer
of Eddington.  In the Observatories he occupied what had been the
famous Eddington's office with a curved door. Our conversations were
usually there, and sometimes, weather permitted, in front of the main
building where Einstein and Eddington took the photograph.

I thank António Luciano Videira, my MSc thesis supervisor in PUC-Rio
de Janeiro, for allowing me to study what I wanted: black holes; and
also through reading, studying and conversation, putting me in direct
connection with Wheeler's tradition, and thus with Einstein and
Oppenheimer, as well as with the works of Penrose, Hawking, and
Carter.

In July 11, 1991, there was an eclipse that would
end in 
Amazonia, Brazil, after passing through the Pacific Ocean, Hawaii,
Mexico, Guatemala, Nicaragua, Costa Rica, Panama, 
and Colombia. I was in my
first job in Observatorio Nacional, Rio de Janeiro, and being a
general relativist thought opportune to test general relativity once
again through light deflection. I start some demarches but soon I got
stuck. I could not find financial support and several people, 
notably Antares Kleber of Observatorio Nacional,
convinced
me that it was a too risky enterprise, as the probability of rain in
Amazonia was high, the eclipse would be in the end of the afternoon, and
it was too difficult to find good instruments to be carried to
the site, CCDs were
not yet around in the corner shop. So I did not go and did not thought
about it again until today.

I thank Funda\c c\~ao para a Ci\^encia e Tecnologia (FCT), Portugal,
for financial support through Grant~No.~UID/FIS/00099/2019.

\end{acknowledgments}


\vskip 1cm


\begin{thebibliography}{99}



\bibitem{einstein1907} 
A. Einstein, ``Über das Relativitätsprinzip und die aus demselben
gezogene Folgerungen'', Jahrbuch der Radioaktivität und Elektronik {\bf 4},
411 (1907); translation, A. Einstein, ``On the relativity principle and 
the conclusions
drawn from it'', in {\it The Collected Papers of Albert Einstein,
Volume 2: The Swiss Years: Writings, 1900-1909}, editors A. Beck and
Peter Havas (Princeton University Press, New Jersey, 1989),
p.~252.


\bibitem{einstein1911}
A. Einstein, ``Über den Einfluß der Schwerkraft auf die Ausbreitung
des Lichtes'', Annalen der Physik {\bf 35}, 898 (1911); translation, A. Einstein, 
``On the
influence of gravitation on the propagation of light'', in
A. Einstein, H. A. Lorentz, H. Minkowski, and H. Weyl, {\it The
Principle of Relativity}, (Methuen and Company, London, 1923), p.~99;
and ``On the influence of gravitation on the propagation of light'',
in {\it The Collected Papers of Albert Einstein, Volume 2: The Swiss
Years: Writings, 1909-1911}, editors A. Beck and D. Howard (Princeton
University Press, New Jersey, 1994), p.~379.


\bibitem{einsteingrossmann1914}
A. Einstein and M. Grossmann,
``Entwurf einer verallgemeinerten Relativit\"atstheorie
und einer Theorie der Gravitation'',
Zeitschrift für Mathematik und Physik
{\bf 62}, 225 (1914);
translation, 
A. Einstein and M. Grossmann, 
``Outline of a generalized theory of relativity
and of a theory of gravitation'',
in {\it The Collected Papers of Albert Einstein, Volume 4: The Swiss
Years: Writings, 1912-1914}, editors A. Beck and D. Howard (Princeton
University Press, New Jersey, 1996), p.~151.










\bibitem{einstein1914aentwurfdeflection}
A. Einstein, ``Physikalische Grundlagen einer
Gravitationstheorie'', 
Naturforschende Gesellschaft in Zürich,
Vierteljahrsschrift 
{\bf 58}, 284 (1914); 
translation, 
A. Einstein, 
``Physical  foundations of a theory
of gravitation'',
in {\it The Collected Papers of Albert Einstein, Volume 4: The Swiss
Years: Writings, 1912-1914}, editors A. Beck and D. Howard (Princeton
University Press, New Jersey, 1996), p.~192.





\bibitem{einstein1914b}
A. Einstein, ``Zur Theorie der Gravitation'',
Naturforschende Gesellschaft in Zürich
Vierteljahrsschrift 
{\bf 59}, IV (1914); 
translation, 
A. Einstein, 
``On the theory of gravitation'',
in {\it The Collected Papers of Albert Einstein, Volume 4: The Swiss
Years: Writings, 1912-1914}, editors A. Beck and D. Howard (Princeton
University Press, New Jersey, 1996), p.~291.


\bibitem{einstein1915perihelion} 
A. Einstein, ``Erklärung der Perihelbewegung des Merkur aus der
allgemeinen Relativitätstheorie'', Sitzungsberichte der Königlich
Preußischen Akademie der Wissenschaften (Berlin), 831 (1915);
translation, A. Einstein,
``Explanation of the perihelion motion of mercury from the
general theory of relativity'', in {\it The collected papers of Albert
Einstein, Volume 6: The Berlin Years: Writings, 1914-1917}, editors
A. Engel and E. Schucking (Princeton University Press, New Jersey,
1997), p.~112.




\bibitem{einstein1915} 
A. Einstein, ``Die Feldgleichungen der Gravitation'', Sitzungsberichte
der Königlich Preußischen Akademie der Wissenschaften (Berlin), 844
(1915); translation, A. Einstein, 
``The field equations of gravitation'',
in {\it The Collected Papers of Albert Einstein,
Volume 6: The Berlin Years: Writings, 1914-1917}, editors A. Engel and
E. Schucking (Princeton University Press, New Jersey, 1997), p.~117.








\bibitem{einstein1916} 
A. Einstein, ``Die Grundlage der allgemeinen Relativitätstheorie'',
Annalen der Physik {\bf 49}, 769 (1916); translation, A. Einstein, ``The
foundations of the general theory of relativity'', in A. Einstein,
H. A. Lorentz,
H. Minkowski, and H. Weyl, {\it The Principle of Relativity}, (Methuen
and Company, London, 1923), p.~109; and A. Einstein, ``The foundations of the
general theory of relativity'' in {\it The collected papers of Albert
Einstein, Volume 6: The Berlin Years: Writings, 1914-1917}, editors
A. Engel and E. Schucking (Princeton University Press, New Jersey,
1997), p.~146.


\bibitem{renn}
J. Renn and M. Schemmel (editors),
{\it The Genesis of General Relativity},
Volumes 1-4, (Springer, Dordrecht, 2007).

\bibitem{clarkebio} R. W. Clark,
{\it Einstein, the Life and Times} 
(Avon Books, New York, 1971).




\bibitem{pais1}
A. Pais,
{\it Subtle  is the Lord: The Science and the Life of
Albert Einstein} (Oxford University Press, Oxford, 1982).


\bibitem{pais2}
A. Pais,
{\it Einstein Lived Here}
(Oxford University Press, Oxford, 1994).



\bibitem{eddingtonondyson1940}
A. S. Eddington,
``Sir Frank Watson Dyson'',
Biographical Memoirs
of Fellows of the Royal Society
{\bf 3}, 159 (1940).


\bibitem{allie}
A. Vibert Douglas,
{\it Arthur Stanley Eddington}
(Thomas Nelson and Sons, London, 1956).


\bibitem{eddbiographykillmister}
C. W. Killmister,
{\it Eddington's Search for a Fundamental Theory:
A Key to the Universe}
(Cambridge University Press, Cambridge, 1994).

\bibitem{eddbiographystanley}
M. Stanley,
{\it Practical Mystic: Religion, Science, and A. S. Eddington}
(University of Chicago Press, Chicago, 2007).

\bibitem{eddingtononcottigham}
A. S. Eddington,
``Forty years of astronomy'',
in {\it Background to Modern Science},
editors J. Needham and W. Pagel
(Cambridge University Press, Cambridge, 1940), p.~117.


\bibitem{bargerolsson}
V. D. Barger and M. G. Olsson,
{\it Classical Mechanics, a Modern Perspective}
(McGraw-Hill, New York, 1973, second edition 1995).


\bibitem{littmann}
M. Littmann, F. Espenak, and 
K. Willcox,
{\it Totality:
Eclipses of the Sun}
(Oxford University Press, Oxford, 2008).




\bibitem{gren}
T. Nordgren,
{\it Sun, Moon, Earth: The History of Solar Eclipses from Omens 
of Doom to Einstein and Exoplanets}
(Basic Books, Philadelphia, 2016).


\bibitem{bauer}
L. A. Bauer, 
``Resumé of observations concerning the solar
eclipse of May 6.9,1919, and the Einstein
effect'',
Science  {\bf 51}, 301  (1920).

\bibitem{soldner}
J. G. von Soldner, ``Über die Ablenkung eines Lichtstrahls von seiner
geradtinigen Bewegung durch die Attraktion eines Weltkörpers'',
Astronomisches Jahrbuch 161 (1801-1804).

\bibitem{jaki}
S. L. Jaki, ``Johann Georg von Soldner and the gravitational bending
of light, with an English translation of his essay on it published in
1801'', Foundations of Physics {\bf 8}, 927 (1978).


\bibitem{cormmach1968}
R. McCormmach, ``John Michell and Henry Cavendish: Weighing the
Stars'', British Journal for the History of Science {\bf 4}, 126
(1968).


\bibitem{nord}
N. A. Doughty, 
{\it Lagrangian Interaction}
(Addison-Wesley, New York, 1990).



\bibitem{perrine1923}
C. D. Perrine,
``Contribution to the history of attempts to test the theory of
relativity by means of astronomical observations'',
Astronomische Nachrichten {\bf 219}, 281 (1923).

\bibitem{stachel}
J. Stachel, ``Eddington and Einstein'', in {\it The Prism of Science},
editor E. Ullmann-Margalit (Boston Studies in the Philosophy of
Science, Boston, 1986), p.~225.


\bibitem{warwick}
A. Warwick, {\it Masters of Theory: Cambridge and the Rise of
Mathematical Physics} (University of Chicago Press, Chicago, 2003).





\bibitem{eddington1912}
A. S. Eddington and C. Davidson, 
``Total eclipse of the sun, 1912 October 10, report on an
expedition to Passa Quatro, Minas Geraes, Brazil'', 
Monthly Notices of the Royal Astronomical Society {\bf 73}, 
386 (1913).


\bibitem{lindemanns}
A. F. Lindemann and F. A. Lindemann, ``Daylight photography of stars
as a means of testing the equivalence postulate in the theory of
relativity'', Monthly Notices of the Royal Astronomical Society {\bf
77}, 140 (1916).




\bibitem{eddington1918}
A. S. Eddington, {\it Report on the Relativity Theory of Gravitation}
(Fleetway Press, London, 1918, second edition 1920).

\bibitem{dyson1919}
F. W. Dyson,
``On the opportunity afforded by the eclipse of 1919 May 29 of
verifying Einstein's theory of gravitation'',
Monthly Notices of the Royal Astronomical Society
{\bf 77}, 445 (1917).


\bibitem{crommelin1919naturebefore}
A. C. D. Crommelin, 
``The eclipse of the sun on May 29'', 
Nature {\bf 102}, 444 (1919). 





\bibitem{eddington1919}
A. S. Eddington, 
``The total eclipse of 1919 May 29 and the
influence of gravitation on light'', Observatory {\bf 42} 119 (1919).




\bibitem{crommelin1919observatory}
A. C. D. Crommelin, 
``The eclipse expedition to Sobral'', 
Observatory {\bf 42}, 368 (1919). 

\bibitem{jjthomson}
J. J. Thomson (chair), 
``Joint Eclipse Meeting
of the Royal Society and
the Royal Astronomical Society'', 
Observatory {\bf 42}, 388 (1919). 






\bibitem{report1920}
F. W. Dyson, A. S. Eddington, and C. Davidson,
``A determination of the deflection of light by the sun's
gravitational field from observations made at the
total eclipse of May 29, 1919'',
Philosophical Transactions of the Royal
Society A {\bf 220}, 291 (1920).




\bibitem{eddstg1920}
A. S. Eddington, {\it Space, Time, and Gravitation: An Outline
of the General Relativity Theory}
(Cambridge University Press, Cambridge, 1920).



\bibitem{whitehead}
A. N. Whitehead, 
{\it Science and the Modern World}
(The Macmillan Company, New York, 1925).

\bibitem{kluber}
H. von Klüber, ``The determination of Einstein's light-deflection in
the gravitational field of the Sun'', Vistas in Astronomy {\bf 3}, 47
(1960).


\bibitem{chandrasekhar}
S. Chandrasekhar,
``Verifying the theory of relativity'',
Notes and Records of the Royal Society of London
{\bf 30}, 249 (1976).



\bibitem{moyer}
D. F. Moyer, 
``Revolution in science: The 1919 eclipse test of general relativity'',
in {\it On the Path of Albert Einstein}, editors A. Perlmutter
and L. F. Scott
(Plenum Press, New York, 1979), p.~55.

\bibitem{earman} 
J. Earman and C. Glymour, 
``Relativity and eclipses: The British eclipse
expeditions of 1919 and their predecessors'',
Historical Studies in the Physical Sciences {\bf 11},
49 (1980).



\bibitem{will2} C. M. Will,
``The 1919 measurement of the deflection of light'',
Classical and Quantum Gravity {\bf 32}, 124001 (2015);
arXiv:1409.7812 [physics.hist-ph].



\bibitem{kennefick}
D. Kennefick,
{\it No Shadow of a Doubt: The 1919 Eclipse that Confirmed 
Einstein's Theory of Relativity}
(Princeton University Press, New  Jersey, 2019).





\bibitem{rrfm1}
R. R. F. Mourão, 
{\it A Teoria da Relatividade}
(Editora Tecnoprint, Rio de Janeiro, 1987).


\bibitem{rrfm2}
R. R. F. Mourão, 
{\it Einstein: de Sobral para o Mundo}
(Editora Universidade Estadual Vale do Acaraú, 
Sobral, 2003).




\bibitem{crispino1}
L. C. B. Crispino and M. C. de Lima, ``Amazonia introduced to general
relativity: The May 29, 1919, solar eclipse from a north-Brazilian
point of view'', Physics in Perspective {\bf 18}, 379 (2016).


\bibitem{crispino2}
L. C. B. Crispino, ``Expeditions for the observation in Sobral, Brazil,
of the May 29, 1919 total solar eclipse'',
International Journal of Modern Physics D
{\bf 27}, 1843004 (2018).


\bibitem{crispino3}
L. C. B. Crispino and M. C. de Lima, ``Expedição norte-americana e
iconografia inédita de Sobral em 1919'', Revista Brasileira de Ensino
de Física {\bf 40}, e1601 (2018).


\bibitem{crispino5}
L. C. B. Crispino and D. J. Kennefick,
``A hundred years of the first experimental test of general relativity'',
Nature Physics {\bf 15}, 416 (2019);
arXiv:1907.10687 [physics.hist-ph].



\bibitem{joycemota}
J. Mota Rodrigues,
{\it Entre Telescópios e Potes De Barro: Expedições Científicas do Eclipse
Solar na Comprovação da Teoria da Relatividade em Sobral-CE 1919}
(Appris Editora, Curitiba, 2019).



\bibitem{crawforscol}
P. Crawford,
``O significado da relatividade no final do século'', 
Colóquio Ciências (Fundação Calouste
Gulbenkian) {\bf 16}, 3 (1995).


\bibitem{simoes1}
P. Crawford and A. Simões, ``O eclipse de 29 de Maio de 1919: A. S.
Eddington e os astrónomos do Observatório da Tapada'', Gazeta
de Física {\bf 32(2,3)}, 22 (2009).

\bibitem{simoes2}
E. Mota, A. Simões, and P. Crawford, ``Einstein in Portugal: Eddington's
expedition to Principe and the reactions of Portuguese
astronomers (1917–25)'', British Journal for the History of
Science {\bf 42}, 245 (2009).

\bibitem{simoes3}
A. Simões, 
``O eclipse de 29 de maio de 1919 e a teoria
da relatividade: um encontro improvável'',
in 
{\it Einstein, Eddington, Eclipse}, editors A. J. S. Fitas,
P. Crawford, and J. P. S. Lemos (Número especial dedicado à exposição
E3 - Einstein Eddington e o Eclipse, Gazeta de Física, Lisbon, 2019),
p.~4.



\bibitem{crawfordgaz}
P. Crawford,
``Einstein e Eddington antes e depois do
eclipse total do Sol de 1919'',
in 
{\it Einstein, Eddington, Eclipse}, editors A. J. S. Fitas,
P. Crawford, and J. P. S. Lemos (Número especial dedicado à exposição
E3 - Einstein Eddington e o Eclipse, Gazeta de Física, Lisbon, 2019),
p.~8.



\bibitem{harvey}
G. M. Harvey,
``Gravitational deflection of light'',
Observatory {\bf 99}, 195 (1979).

\bibitem{hawking}
S. W. Hawking, {\it A Brief History of Time,
from the Big Bang to Black Holes} 
(Bantam Books, New York, 1988).


\bibitem{waymanmurray} 
P. A. Wayman and C. A. Murray,
``Relativistic light deflections'',
Observatory {\bf 109}, 189 (1989).



\bibitem{fitas}
A. J. S. Fitas, ``The Portuguese academic community and the
theory of relativity'', e-Journal of Portuguese History  {\bf 3}, 2 (2005).



\bibitem{jpsl1}
J. P. S. Lemos,
``General relativity, differential geometry, and unitary theories
in the work of Mira Fernandes'',
in
{\it Proceedings of the 12th Marcel Grossmann
Meeting on General Relativity - MG12},
editors R. Jantzen et al, (World Scientific, Singapore,
2012), p.~1745; arXiv:1011.6269 [physics.hist-ph].




\bibitem{jpsl2}
J. P. S. Lemos,
``Unitary theories in the work of Mira Fernandes (beyond general
relativity and differential geometry)'',
in 
{\it Aureliano Mira Fernandes},
editors L. Saraiva and J. T. Pinto
(Numero Especial - Aureliano Mira Fernandes,
Boletim da Sociedade Portuguesa
de Matemática, Lisbon, 2010), p.~147;
arXiv:1012.5093 [physics.hist-ph].




\bibitem{jpsl3}
J. P. S. Lemos, ``A introdução da relatividade em Portugal e Aureliano de
Mira Fernandes", Gazeta de Física {\bf 34(2)}, 27 (2011).





\bibitem{santoslucas} A. Santos Lucas, {\it Lições sobre a Teoria da
Relatividade, Apontamentos de Física-Matemática de António dos Santos
Lucas Compilados por Francisco de Paula Leite Pinto} (Manuscript
edition from the author,
Lisbon, 1922-1923).



\bibitem{ildeu}
I. C. Moreira and A. A. P. Videira (editors), {\it Einstein e o
Brasil} (Editora Universidade Federal Rio de Janeiro, Rio de Janeiro,
1995).


\bibitem{tomasquim}
A. T. Tomasquim, {\it Einstein: o Viajante da Relatividade na América
do Sul} (vieira \& lent casa editora, Rio de Janeiro, 2003).


\bibitem{moyer2}
D. F. Moyer,
``Origins of Dirac's electron''.
American Journal of Physics {\bf 49}, 944 (1981).


\bibitem{os}
J. R. Oppenheimer and H. Snyder,
``On continued gravitational contraction'',
Physical Review  {\bf 56}, 455 (1939).


\bibitem{thornebhs}
K. S. Thorne, {\it Black Holes and Time Warps,
Einstein's Outrageous Legacy}
(W. W. Norton and Company, New York, 1994).

\bibitem{dyson} F. Dyson,
``The scientist as rebel'',
 American Mathematical Monthly
{\bf 103}, 800 (1996).



\bibitem{herdeirolemos}
C. A. R. Herdeiro and  J. P. S. Lemos,
``O buraco negro cinquenta anos depois:
A génese do nome'', Gazeta de Física {\bf 41(2)}, 2
(2018); for an English version see, C. A. R. Herdeiro and  J. P. S. Lemos,
``The black hole fifty years after: Genesis of the name'',
arXiv:1811.06587 [physics.hist-ph].




\bibitem{eisenstaedt}
J. Eisenstaedt,
{\it 
The Curious History of Relativity: How Einstein's Theory 
of Gravity Was Lost and Found Again}
(Princeton University Press, New Jersey, 2006).


\bibitem{israel}
W. Israel, 
``Dark stars: the evolution of an idea'', 
in {\it 300 Years of Gravitation},
editors S. W. Hawking and W. Israel 
(Cambridge University Press, Cambridge,
1987), p.~198.






\bibitem{eddmathematicalrelativity}
A. S. Eddington, {\it The Mathematical Theory of Relativity}
(Cambridge University Press, Cambridge, 1923).

\bibitem{adler}
R. Adler, M. Bazin, and M. Schiffer,
{\it Introduction to General Relativity}
(McGeaw-Hill, New York, 1965, second edition 1975).


\bibitem{weinberg} 
S. Weinberg,
{\it Gravitation and Cosmology: Principles and Applications of the
General Theory of Relativity}
(Wiley, New York, 1972).

\bibitem{mtw}
C. Misner, K. S. Thorne, 
and J. A. Wheeler,
{\it Gravitation}
(Freeeman, San Francisco, 1973).



\bibitem{wald}
R. M. Wald, 
{\it General Relativity}
(University of Chicago Press, Chicago, 1984).



\bibitem{dinverno}
R. d'Inverno,
{\it Introducing Einstein's Relativity}
(Clarendon Press, Oxford, 1992).



\bibitem{krasinski}
J. Pleba\'nski and A. Krasi\'nski,
{\it An Introduction to General Relativity and Cosmology}
(Cambridge University Press, Cambridge, 2006).


\bibitem{ryder}
L. Ryder, 
{\it Introduction to General Relativity}
(Cambridge University Press, Cambridge, 2009).




\bibitem{mauritaniatexas}
R. A. Brune, C. L. Cobb, B. S. DeWitt, C.
DeWitt-Morette, D. S. Evans, J. E. Floyd, B. F.
Jones, R. V. Lazenby, M. Marin, R. A. Matzner,
A. H. Mikesell, M. R. Mikesell, R. I. Mitchell,
M. P. Ryan, H. J. Smith, A. Sy, and C. D.
Thompson, plus the secretarial, fiscal, engineering, and maintenance
staffs of the Department of Astronomy and Center for
Relativity of the University of Texas,
``Gravitational deflection of light: solar eclipse of 30 June 1973 I.
Description of procedures and final results'',
Astronomical Journal {\bf 81}, 452 (1976).




\bibitem{bruns}
D. G. Bruns, ``Gravitational starlight deflection measurements during
the 21 August 2017 total solar eclipse'', Classical and Quantum
Gravity {\bf 35}, 075009 (2018); arXiv:1802.00343 [astro-ph.IM].




\bibitem{schneider}
P. Schneider,
C. Kochanek, and J. Wambsganss,
{\it Gravitational Lensing: Strong, Weak and Micro}
(Springer Verlag, Berlin, 2006). 



\bibitem{tereno}
I. Tereno, 
``100 anos de lentes gravitacionais'', 
in 
{\it Einstein, Eddington, Eclipse}, editors A. J. S. Fitas,
P. Crawford, and J. P. S. Lemos (Número especial dedicado à exposição
E3 - Einstein Eddington e o Eclipse, Gazeta de Física, Lisbon, 2019),
p.~43.



\bibitem{jpslemosherdeirocardosogazeta}
J. P. S. Lemos, C. A. R. Herdeiro, and V. Cardoso,
``Einstein e Eddington e as consequências
da relatividade geral: Buracos negros e
ondas gravitacionais'', 
in 
{\it Einstein, Eddington, Eclipse}, editors A. J. S. Fitas,
P. Crawford, and J. P. S. Lemos (Número especial dedicado à exposição
E3 - Einstein Eddington e o Eclipse, Gazeta de Física, Lisbon, 2019),
p.~36; for an English version see
J. P. S. Lemos, C. A. R. Herdeiro, and V. Cardoso,
```Einstein, Eddington and the consequences of
general relativity: Black holes and gravitational waves'',
arXiv:1910.xxxx [physics.hist-ph].



\bibitem{perryman}
M. Perryman, {\it Astronomical Applications of Astrometry: Ten Years of
    Exploitation of the Hipparcos Satellite Data} (Cambridge
    University Press, Cambridge, 2009).

\bibitem{perrymanpaper}
M. Perryman, D. N. Spergel, and L. Lindegren, ``The Gaia inertial
reference frame and the tilting of the Milky Way disk'', Astrophysical
Journal {\bf 789}, 166 (2014); arXiv:1406.0129 [astro-ph.GA].






\bibitem{shapiro}
I. I. Shapiro, ``New method for the detection of
light deflection by solar gravity'',
Science, {\bf 157}, 806 (1967).


\bibitem{shapiro2}
D. E. Lebach, B. E. Corey, I. I. Shapiro, M. I. Ratner, J. C. Webber,
A. E. E. Rogers, J. L. Davis, and T. A. Herring, 
``Measurement of the solar gravitational deflection of radio waves
using very-long-baseline interferometry'',
Physical Review Letters {\bf 75}, 1439 (1995).



\bibitem{fomalont}
E. Fomalont, S. Kopeikin, G. Lanyi, and J. Benson,
``Progress in measurements of the gravitational bending of
radio waves using the VLBA'',
Astrophysical Journal {\bf 699}, 1395 (2009);
arXiv:0904.3992 [astro-ph.CO].



\bibitem{sazhin}
M. V. Sazhin, V. E. Zharov, T. A. Kalinina, and V. N. Sementsov,
``Cosmology and astrometry'',
Astronomy Reports {\bf 62}, 1026 (2018).



\bibitem{will1}
C. M. Will, 
``The confrontation between general relativity and experiment'',
Living Reviews in Relativity {\bf 17}, 4 (2014);
arXiv:1403.7377 [gr-qc].


\bibitem{ildeugazeta}
I. C. Moreira, 
``O eclipse Solar de 1919 e as atividades
comemorativas no Brasil'',
in 
{\it Einstein, Eddington, Eclipse}, editors A. J. S. Fitas,
P. Crawford, and J. P. S. Lemos (Número especial dedicado à exposição
E3 - Einstein Eddington e o Eclipse, Gazeta de Física, Lisbon, 2019),
p.~32.


\bibitem{latas}
J. Latas, ``A pedra fundamental de um legado no
Príncipe, 100 anos depois'',
in 
{\it Einstein, Eddington, Eclipse}, editors A. J. S. Fitas,
P. Crawford, and J. P. S. Lemos (Número especial dedicado à exposição
E3 - Einstein Eddington e o Eclipse, Gazeta de Física, Lisbon, 2019),
p.~30.




\bibitem{jpslfitascrawford}
A. J. S. Fitas, P. Crawford, and J. P. S. Lemos (editors), {\it Einstein,
Eddington, Eclipse} (Número especial dedicado à exposição E3 -
Einstein Eddington e o Eclipse, Gazeta de Física, Lisbon, 2019).


\end{thebibliography}
\end{document}